\documentclass[12pt]{JHEP3}

\usepackage{epsfig,subfigure}

\newcommand{\bmat}{\left(\begin{array}}
\newcommand{\emat}{\end{array}\right)}

\def\yzero{\smash{\hbox{$y\kern-4pt\raise1pt\hbox{${}^\circ$}$}}}

\def\beq{\begin{equation}}
\def\eeq{\end{equation}}
\def\beqa{\begin{eqnarray}}
\def\eeqa{\end{eqnarray}}

\def\-{\hphantom{-}}
\def\ov{\overline}
\def\s2{\frac{1}{2}}

\def\beq{\begin{equation}}
\def\eeq{\end{equation}}
\def\beqa{\begin{eqnarray}}
\def\eeqa{\end{eqnarray}}

\def\Tr{{\rm Tr \,}}
\def\diag{{\rm diag \,}}

\def\IF{\relax{\rm I\kern-.18em F}}
\def\II{\relax{\rm I\kern-.18em I}}
\def\IP{\relax{\rm I\kern-.18em P}}

\def\cp{{\cal P}}
\def\IC{\bf C}
\def\IZ{\bf Z}
\def\IS{\bf S}
\def\IM{\bf M}
\def\IR{\bf R}

\def\id{{\bf 1}}

\def\NN{{\cal N}}
\def\Dsl{\,\raise.15ex\hbox{/}\mkern-13.5mu D} 
\def\IT{\bf T}
\def \one{\relax{\rm 1\kern-.26em I}}
 \def\cp#1{\relax\ifmmode {\IP\kern-2pt{}_{#1}}\else $\IP\kern-2pt{}_{#1}$\=fi}

\catcode`\@=11
\newdimen\@rotdimen
\newbox\@rotbox

\def\@vspec#1{\special{ps:#1}}
\def\@rotstart#1{\@vspec{gsave currentpoint currentpoint translate
   #1 neg exch neg exch translate}}
\def\@rotfinish{\@vspec{currentpoint grestore moveto}}
\def\@rotr#1{\@rotdimen=\ht#1\advance\@rotdimen by\dp#1%
   \hbox to\@rotdimen{\hskip\ht#1\vbox to\wd#1{\@rotstart{90 rotate}%
   \box#1\vss}\hss}\@rotfinish}
\def\@rotl#1{\@rotdimen=\ht#1\advance\@rotdimen by\dp#1%
   \hbox to\@rotdimen{\vbox to\wd#1{\vskip\wd#1\@rotstart{270 rotate}%
   \box#1\vss}\hss}\@rotfinish}%
\def\@rotu#1{\@rotdimen=\ht#1\advance\@rotdimen by\dp#1%
   \hbox to\wd#1{\hskip\wd#1\vbox to\@rotdimen{\vskip\@rotdimen
   \@rotstart{-1 dup scale}\box#1\vss}\hss}\@rotfinish}%
\def\@rotf#1{\hbox to\wd#1{\hskip\wd#1\@rotstart{-1 1 scale}%
   \box#1\hss}\@rotfinish}%
\def\rotate{\@ifnextchar[{\@rotate}{\@rotate[l]}}
\def\@rotate[#1]#2{\setbox\@rotbox=\hbox{#2}\@nameuse{@rot#1}\@rotbox}

\catcode`\@=12

\title{Realistic D-Brane Models on Warped Throats: Fluxes,
  Hierarchies and Moduli Stabilization}

\author{J.F.G. Cascales $^1$, M.P. Garc\'{\i}a del Moral $^1$,
  F. Quevedo $^2$, A. M. Uranga $^1$  
\\
$^1$ Departamento de F\'{\i}sica Te\'orica C-XI,  and 
     Instituto de F\'{\i}sica Te\'orica C-XVI\\
     Universidad Aut\'onoma de Madrid
     Cantoblanco, 28049 Madrid Spain.\\
$^2$ DAMTP, Centre for Mathematical Sciences\\
               University of Cambridge,
               Cambridge CB3 0WA UK.}

\abstract{
We describe the construction of string theory models with semirealistic 
spectrum in a sector of (anti) D3-branes located at an orbifold 
singularity at the bottom of a highly warped throat geometry, which 
is a generalisation of the Klebanov-Strassler deformed conifold. 
These models realise the Randall-Sundrum proposal to naturally generate 
the Planck/electroweak hierarchy in a concrete string theory embedding, and
yielding interesting chiral open string spectra. We describe examples with
Standard Model gauge group (or left-right symmetric extensions) and three
families of SM fermions, with correct quantum numbers including hypercharge.
The dilaton and complex structure moduli of the geometry are stabilised by 
the 3-form fluxes required to build the throat. We describe diverse issues
concerning the stabilisation of geometric K\"ahler moduli, like blow-up 
modes of the orbifold singularities, via D term potentials and 
gauge theory non-perturbative effects, like gaugino condensation. 
This local geometry, once embedded in a full compactification, could give 
rise to models with all moduli stabilised, and with the potential to 
lead to de Sitter vacua. Issues of gauge unification, proton stability, 
supersymmetry breaking and Yukawa couplings are also discussed.
}

 \preprint{}

\preprint{DAMTP-2003-133\\  IFT-UAM/CSIC-03-50 \\FTUAM-03-28 \\ hep-th/0312051}

\begin{document}

\makeatletter \@addtoreset{equation}{section} \makeatother
\renewcommand{\theequation}{\thesection.\arabic{equation}}


\setcounter{page}{1} \pagestyle{plain}
\renewcommand{\thefootnote}{\arabic{footnote}}
\setcounter{footnote}{0}


\section{Introduction}

Important progress has been made during the past few years regarding 
string theory model building. Chiral models resembling very much the
structure of the standard model of particle physics have been explicitly  
constructed from type II and type I compactifications with D-branes 
(using e.g. D-branes at singularities and intersecting D-branes, see e.g. 
\cite{reviews} for reviews).
 These models however leave open the question 
of moduli stabilisation. The few supersymmetric models explicitly 
constructed have not fully realistic spectra, and in addition have many 
moduli fields with flat potentials. The non-supersymmetric models lead to
spectra closer to the Standard Model, but there is no good control of the 
scalar field potentials, and in  most models they render the models 
unstable towards runaway zero-coupling or decompactification limits. A 
related further problem is that the smallness of electroweak scale, 
usually associated to a large size of the extra dimensions, is left 
unexplained in these models.

On the other hand, progress has been recently achieved in 
constructing non-realistic models with stabilisation 
of the dilaton and many geometric moduli via the introduction of a background 
of NSNS and RR field strength fluxes \cite{fluxes,gvw,gkp} 
 \footnote{See \cite{blt,cascur} for 
compactifications with fluxes and semirealistic spectra.}(in some cases, 
the remaining moduli may be stabilised via non-perturbatively generated 
superpotentials \cite{kklt,egq}). Furthermore such flux compactifications 
naturally involve a string theory realisation of the Randall-Sundrum (RS) 
scenario \cite{rs}, since they may lead to a hierarchy of scales by means 
of a warp factor in the higher dimensional metric, sourced by the energy 
carried by the flux background \cite{verlinde,gkp}. 

A concrete construction reproducing the key features of the RS setup is 
provided by considering flux compactifications including a Klebanov-Strassler 
(KS) throat \cite{ks}. Namely, one considers compactification on Calabi-Yau 
spaces containing deformed conifold singularities \cite{px}, and 
introduces a large flux on the corresponding 3-sphere (and its dual 
cycle). The flux back-reaction on the metric creates a long throat, which 
ends smoothly at a tip containing the original 3-sphere supporting the 
flux. Since 3-form fluxes gravitate and carry 4-form charge, they create 
a supergravity background of the black 3-brane form, and the structure of 
the central region of the throat is an slice of anti-de Sitter space, 
AdS$_5$. The throat is cut off by the 3-sphere at the strongly warped 
end, and by the original Calabi-Yau compact space at the weakly warped 
end. These regions correspond to the infrared and ultraviolet branes in 
the RS construction \cite{verlinde, gkp}.

This is a very appealing scenario that includes several desirable properties,
such as moduli fixing, natural hierarchy and de Sitter vacua. However there is
no concrete proposal on how to include the Standard Model (or a gauge 
sector similar to it) in the scenario. This is the task we undertake in 
the present paper.

\begin{figure}
\begin{center}
\centering
\epsfysize=3.5in
\leavevmode
\epsfbox{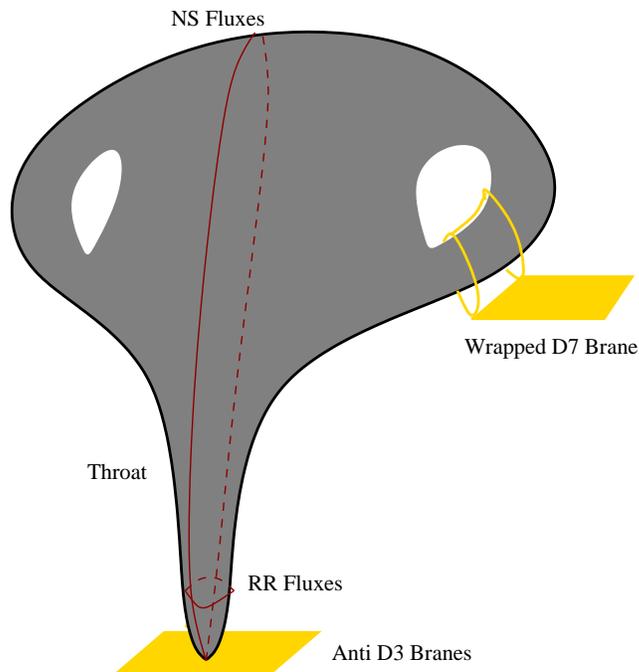}
\end{center}
\caption[]{\small Description of a deformed conifold with 3-form fluxes 
(a KS throat) embedded in a compact geometry, with anti-D3-branes trapped 
at the tip of the throat. Beyond the throat, the compactifications may 
include other ingredients, like D7-branes wrapped on 4-cycles, etc, which 
are not relevant for the generation of the warp factor on the throat, 
but may lead to other interesting effects (like non-perturbative 
superpotentials).} 
\label{throatCY}
\end{figure}

In order to generate the correct weak scale via the RS hierarchy, the Standard 
Model should be localised at the tip of the throat. A possibility would be to 
localise it on the world-volume of a set of D3-branes. However, moduli of 
D3-branes in imaginary self-dual fluxes of the kind discussed 
in \cite{gkp} are not lifted. 
Hence one would have to face the problem of fixing these extra moduli, 
in order to explain the stabilisation of the D3-branes at the tip of 
the throat. An interesting alternative is to consider the SM to be
localised
 on anti-D3-branes (denoted $\ov{\rm{D3}}$-branes in what follows). These 
objects 
appear naturally in the construction of de Sitter vacua of string theory 
using flux stabilisation of moduli, in particular their tension is 
responsible for the lifting of the vacuum energy to a positive value. In 
the setup including KS throats, anti-D3-branes are interesting because they 
are attracted by the fluxes and naturally fall to the tip of the throat 
\cite{kpv}. 
Hence they are the natural objects to localise the Standard Model on 
their world-volume, and hence at the tip of the throat. A picture of the 
configuration is shown in figure \ref{throatCY}. In the present paper we consider both possibilities.

Clearly, to obtain the Standard Model it is not enough to have a stack of 
D-branes, one needs the world-volume spectrum to lead to chiral fermions, 
with correct quantum numbers, etc. For D3-, $\ov{\rm D3}$-branes, the massless 
world-volume spectrum depends basically on the local geometry around them. 
It is then natural to apply previous results of D-brane model building to 
the present problem, in particular the bottom-up approach in \cite{aiqu} 
(see also \cite{leigh}).

In \cite{aiqu} a new approach to build string models was proposed. 
The idea is to exploit the `modular' structure for the string models 
constructed with D-branes (in other words, exploit locality in 
the extra dimensions). Namely, one first tailors a local D-brane  
configuration that 
can accommodate the Standard Model, and subsequently embeds it into 
different possible compactification manifolds. This approach separates the 
local properties of the models, such as the gauge group, the massless 
matter spectrum, running of gauge coupling, etc,  from properties 
depending strongly on the global features of the compactification, such 
as supersymmetry breaking, scalar field potentials, etc. 

A large class of local D-brane configurations leading to chiral 4d  
world-volume gauge sectors is provided by D3-branes (or 
$\ov{\rm D3}$-branes) at singularities. It is thus natural to combine techniques of model building with $\ov{\rm{D3}}$-branes at singularities with the construction of 
highly warped throats using deformed conifolds with fluxes. Indeed in this 
paper we construct explicit geometries containing deformed conifolds, and 
orbifold singularities sitting at the corresponding 3-spheres. Introduction of 
an explicit set of suitable 3-form fluxes leads to a warped throat, with 
the compact 3-cycles and the orbifold singularity at its tip. Finally 
introducing a set of $\ov{\rm{D3}}$-branes and D7-branes (all dynamically 
trapped at the tip of the throat) at the orbifold singularity, we obtain 
semirealistic chiral gauge sectors localised at the infrared end of the 
throat, so that their natural scale is strongly red-shifted with respect 
to the fundamental scale of the theory.

The paper is organised as follows. The next two sections are
background material. In section 2 we briefly review the construction of 
chiral string models from D-branes at singularities. In section 3 we 
describe the introduction of fluxes in type IIB string theory 
compactifications, and their effect in fixing the dilaton and complex 
structure moduli, and in providing the string theory realisation of 
the Randall-Sundrum hierarchy via a strongly warped throat,
based on a deformed conifold background (fluxes on a compact 3-sphere).

In section 4 we provide the explicit construction of a geometry containing a
set of compact 3-spheres and $\IC^3/\IZ_3$ orbifold singularities. Sets of 
D-branes located at the latter naturally lead to chiral gauge sectors on 
their world-volume. We point out that the construction, involving a double 
elliptic fibration, naturally includes a $\IT^4$ at the tip of the throat 
were D7 branes can wrap around. This allows to enrich the D-brane 
configurations at the tip of the throat, and improve the resulting chiral 
models. We also describe the introduction of a particular set of 
3-form fluxes in this background, and the resulting stabilisation of 
complex structure moduli. These fluxes create a KS-like throat, with the 
D-brane configuration located at its tip.

In section 5 we present explicit examples of realistic models with the 
Standard Model gauge group or its left-right extensions. We present two 
classes of models depending on whether the Standard Model is realized on 
a stack of D3- or $\ov{\rm D3}$-branes. Section 6 is devoted to a 
discussion on how the extra moduli necessarily introduced by our 
construction can be fixed either by D-term potentials and/or 
nonperturbative effects. Finally, in section 7 we discuss some 
phenomenological properties of the models. We finish with some general 
discussions and open questions. Appendix A contains a technical point on 
the consistency of the set of fluxes we introduce in our geometry. A 
second appendix contains several tables with explicit massless spectrum 
of some of the most relevant models discussed in section 5.

\section{D-Branes at Singularities}

Let us briefly review the bottom-up approach to phenomenological model 
building with D-branes at singularities. We centre our description on 
$\ov{\rm D3}$-branes at singularities. A more general and comprehensive 
discussion, centred on D3-branes, can be obtained from \cite{aiqu}.

\begin{figure}
\begin{center}
\centering
\epsfysize=2.5in
\leavevmode
\epsfbox{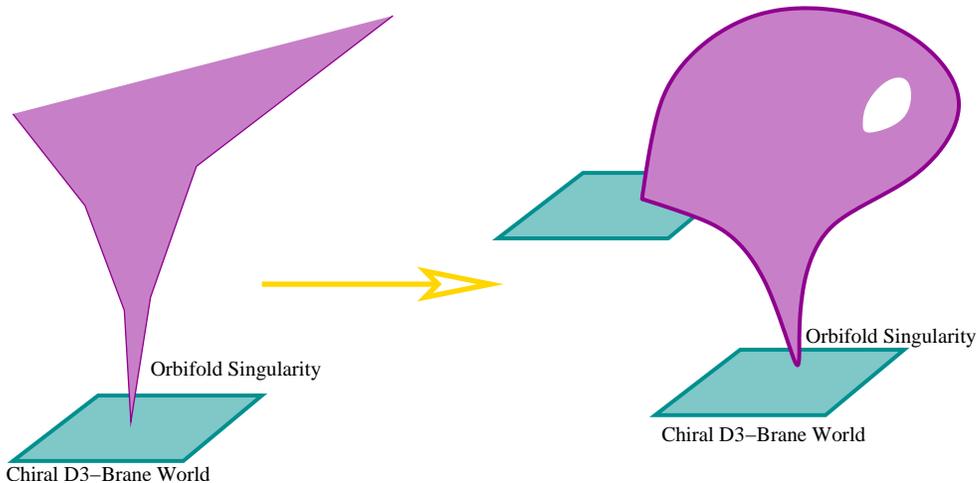}
\end{center}
\caption[]{\small An illustration of the bottom-up approach. D3-branes
at an orbifold singularity provide examples of chiral theories in
which the Standard Model can be embedded. Most of the properties of
the model depend on the structure of the singularity. This local model can
then be embedded in many different string compactifications, as long as
the compact manifold has the same type of orbifold singularity.}
\label{bottomup}
\end{figure}       

For concreteness we centre on $\IC^3/\IZ_N$ orbifold singularities, with 
$\IZ_N$ generated by the action
\beq
(z_1, z_2, z_3)\rightarrow \left( \alpha^{l_1} z_1, \alpha^{l_2} z_2,
\alpha^{l_3} z_3 \right)
\label{orbifoldaction}
\eeq
with $\alpha = e^{2\pi i /N}$. For $l_1+l_2+l_3=0$ mod $N$ the orbifold
preserves an ${\cal N}=2$ supersymmetry in the bulk, further reduced to $\NN=1,0$ when D-branes are introduced. We restrict to this
situation, since non-supersymmetric orbifolds contain closed string 
tachyons in twisted sectors, which complicate the system.

Let us introduce a set of $n$ $\ov{\rm{D3}}$-branes spanning $\IM_4$ and 
sitting at the singular point in $\IC^3/\IZ_N$. Before the orbifold 
projection, the set of $\ov{\rm{D3}}$-branes in flat space leads to a 
$\NN=4$ $U(n)$ world-volume gauge theory. Decomposing it with respect to 
the $\NN=1$ supersymmetry preserved by the branes, we have a $U(n)$ 
vector multiplet $V$, and three chiral multiplets $\Phi_a$ in the adjoint 
representation. The world-volume gauge field theory for 
$\ov{\rm{D3}}$-branes 
at the orbifold geometry is obtained \cite{dm} from the above one by 
keeping the $\IZ_N$-invariant states. The $\IZ_N$ has a geometric action 
on $\Phi_a$, similar to (\ref{orbifoldaction}), and an action on the 
gauge degrees of freedom, given by conjugation of the Chan-Paton 
wavefuncion $\lambda$ by an $n\times n$ unitary matrix 
$\gamma_{\theta,\bar{3}}$ or order $N$. 
\beq
\lambda\rightarrow \gamma_{\theta,\bar{3}}\, \lambda \,
\gamma_{\theta,\bar{3}}^{-1}
\eeq
Without loss of generality, $\gamma_{\theta,\bar{3}}$ can be 
diagonalised, to take the simple form
\beqa
\gamma_{\theta,\bar{3}} = \diag\left(\id_{n_0}, \alpha \id_{n_1},\cdots,
  \alpha^{N-1} \id_{n_{N-1}}\right)
\label{cpone}
\eeqa
Here $\id_{n_k}$ is the identity matrix in $n_k$ dimensions, and integers 
$n_k$ satisfy $\sum_k n_k = n$.

Imposing invariance under the combined geometric and Chan-Paton $\IZ_N$ 
actions, we have the projections
\beqa
V \, : \, \lambda = \gamma_{\theta,\bar{3}} \, \lambda\, 
\gamma_{\theta,\bar{3}}^{-1} 
\quad ; \quad 
\Phi_a \, : \, \lambda = \alpha^{l_a}\, \gamma_{\theta,\bar{3}}\, \lambda 
\, \gamma_{\theta,\bar{3}}^{-1} 
\eeqa
The resulting spectrum of $\NN=1$ multiplets is
\beqa
&\NN=1 \,\, {\rm Vect. Mult.} \quad & U(n_0)\times U(n_1)\times \cdots 
\times U(n_{N-1}) \nonumber \\
& \NN=1 \,\, {\rm Ch.Mult} \quad &
\sum_{a=1}^3\sum_{i=0}^{N-1}\left({\bf n}_i, {\ov{\bf n}}_{i+l_a}\right)
\label{aqui}
\eeqa
with the index $i$ defined modulo $N$. 
This kind of spectrum is usually encoded in quiver diagrams, where gauge 
factors are represented by nodes, and bi-fundamental fields $(n_i,{\ov 
n}_j)$ are represented by oriented arrows from the $i^{th}$ to the 
$j^{th}$ node, see below.
 
From (\ref{aqui}) we can extract a very simple but powerful conclusion:
Only for the $\IZ_3$ orbifold, $(l_1,l_2,l_3)=(1,1,-2)$, will we get a 
matter spectrum arranged in three identical copies or families. Indeed,
only for that case we have $l_1=l_2=l_3$ mod $N$. 
Therefore, and quite remarkably, the maximum number of families for this 
class of models is three, and it is obtained for a unique twist, the 
$\IZ_3$ twist. 

It is important to realise that a consistent configurations of D-branes should obey cancellation of tadpoles for RR fields with compact support in 
the transverse space. This in particular guarantees cancellation of 
non-abelian anomalies, and the cancellation of mixed $U(1)$ anomalies via 
a Green-Schwarz mechanism \cite{gsorbifold}. If only $\ov{\rm{D3}}$-branes 
are present 
at the orbifold singularity, the RR tadpole cancellation condition reads
\beqa
\prod_{a=1}^3 \,2\,\sin(\pi kl_a/N) \, \Tr \gamma_{\theta^k,\bar{3}}\, =\, 
0
\eeqa
For $\IZ_3$ this implies $n_0=n_1=n_2$, which only allows for 
phenomenologically unattractive possibilities (and does not allow for 
more promising ones e.g. $n_0=3, n_1=2, n_2=1$).

This can be improved by enriching the configurations, introducing other 
kinds of D-branes passing through the singularity. We will centre on 
D7-branes spanning two complex planes in the six extra dimensions, which 
are a natural ingredient in IIB flux compactifications \cite{gkp}. For 
concreteness, we centre on a stack of $w$ D7$_3$-branes, spanning the 
submanifold $z_3=0$ in the geometry, hence transverse to the third 
complex plane, which we assume has even $l_3$. 

Hence, in addition to the $\bar{3}\bar{3}$ spectrum, there is a sector of 
$\bar{3}7+7\bar{3}$ open strings \footnote{The 77 open string sector is 
more dependent on global features of the compactification. We skip it for 
the moment, but include it explicitly in our examples.}. Before the 
orbifold projection, it leads to a 4d $\NN=2$ hypermultiplet transforming 
in the bi-fundamental representation $(n,\bar{w})$. In order to quotient 
by the orbifold action, one should introduce an $w\times w$ Chan-Paton 
matrix $\gamma_{\theta,7_3}$ implementing it in the D7-brane gauge 
degrees of freedom
\beqa
\gamma_{\theta,7_3}=\diag\left(\id_{w_0}, \alpha \id_{w_1},\cdots,
  \alpha^{N-1} \id_{w_{N-1}}\right)
\eeqa
In addition we have the geometric action of $\IZ_N$ on the 
$\bar{3}7+7\bar{3}$ fields. Since the orbifold breaks the supersymmetry 
preserved by the D7- and the $\ov{\rm{D3}}$-branes, scalars and fermions 
pick up different geometric phases. The projection conditions read \beqa
{\rm Cmplx.Scalar} :  \lambda_{\bar{3}7} = 
\gamma_{\theta,\bar{3}} \lambda_{\bar{3}7} \gamma_{\theta,7_3}^{-1}; \quad 
{\rm 4d}\,\,{\rm Weyl} \,{\rm Ferm.}  :  \lambda_{\bar{3}7} = 
e^{-i\pi l_3/N}
\gamma_{\theta,\bar{3}} \lambda_{\bar{3}7} \gamma_{\theta,7_3}^{-1}   
\eeqa
The resulting spectrum is
\beqa
& {\rm Cmplx.Scalars} \quad &
\sum_{i=0}^{N-1} \left[ \,\left( {\bf n}_i, \bar{\bf w}_{i} \right)
+  \left( \bar{\bf n}_i, \bf w_{i} \right)\, \right]  \nonumber\\
& {\rm 4d}\,\,{\rm Weyl} \,{\rm Ferm.} \quad &
\sum_{i=0}^{N-1} \left[ \, \left( \bar{\bf n}_i, \bf w_{i-\frac 12 l_3} 
\right) +  
\left( {\bf n}_{i-\frac 12 l_3}, \bar{\bf w}_{i} \right)
\, \right] 
\eeqa
The RR tadpole cancellation condition reads
\beqa
\prod_{a=1}^3 \, 2\sin(\pi kl_a/N) \, \Tr \gamma_{\theta^k,\bar{3}}\, 
-2\sin(\pi kl_3/N) \, \Tr \gamma_{\theta^k,7_3}\, =\, 0
\eeqa

For the interesting case of $\IZ_3$, the full $\bar{3}\bar{3}$ and 
$\bar{3}7+7\bar{3}$ spectrum on the $\ov{\rm D3}$-brane world-volume is
\beqa
\bar{3}\bar{3} \quad & \quad \NN=1 \,\, {\rm Vect. Mult.} \quad & 
U(n_0)\times U(n_1)\times U(n_2) \nonumber \\
& \NN=1 \,\, {\rm Ch.Mult} \quad &
3\, [\, (n_0,\bar{n}_1)\, +\, (n_1,\bar{n}_2)\, + \, (n_2,\bar{n}_0)\, ]
\\
\bar{3}7+7\bar{3} & {\rm Cmplx.Scalars} &
[ \, ( n_0, \bar{w}_0) +  ( n_1, \bar{w}_1) +  ( n_2, \bar{w}_2) +  
( \bar{n}_0, w_0) +  ( \bar{n}_1, w_1) +  ( \bar{n}_2, w_2) \, ] \nonumber
\\
& {\rm 4d}\,\,{\rm Weyl} \,{\rm Ferm.} &
[ \, ( n_0, \bar{w}_2) +  ( n_1, \bar{w}_0) +  ( n_2, \bar{w}_1) +  
( \bar{n}_2, w_0) +  ( \bar{n}_0, w_1) +  ( \bar{n}_1, w_2) \, ] \nonumber
\eeqa

Let us finally provide the spectrum for a system of D3- and D7-branes.
Following \cite{aiqu} we can consider for a configuration of D3-branes 
(with CP matrix $\gamma_{\theta,3}$ given by (\ref{cpone}) and D7-branes, 
as above. The resulting $\NN=1$ supersymmetric spectrum is
\beqa
33 \quad & \quad \NN=1 \,\, {\rm Vect. Mult.} \quad & 
U(n_0)\times U(n_1)\times U(n_2) \nonumber \\
& \NN=1 \,\, {\rm Ch.Mult} \quad &
3\, [\, (n_0,\bar{n}_1)\, +\, (n_1,\bar{n}_2)\, + \, (n_2,\bar{n}_0)\, ]
\\
37+73 & \NN=1 \,\, {\rm Ch.Mult} &
[ \, ( n_0, \bar{w}_1) +  ( n_1, \bar{w}_2) +  ( n_2, \bar{w}_0) +  
(\bar{n}_0, w_2) +  ( \bar{n}_1, w_0) +  ( \bar{n}_2, w_1) \, ] \nonumber
\eeqa

\begin{figure}
\begin{center}
\centering
\epsfysize=3.9cm
\leavevmode
\epsfbox{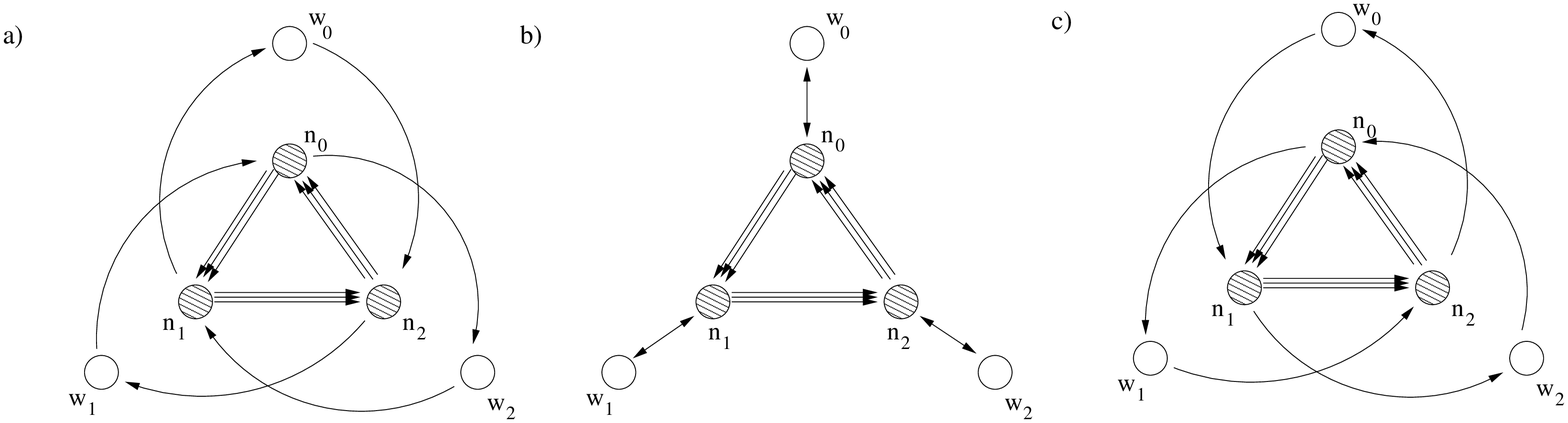}
\end{center}
\caption[]{\small This is the quiver diagram corresponding to a general $\IC^3 /\IZ_3$ model. In the non-SUSY case (models with $\ov{\textrm{D3}}$ and D7-branes), fermions and scalars transform in different representations of the gauge group, namely those displayed (respectively) in \textbf{a)} and \textbf{b)}.For the supersymmetric case, both scalars and fermions fill in chiral multiplets of the preserved SUSY, and transform in the bifundamentals displayed in \textbf{c)}.}
\label{quivergen}
\end{figure}

This general spectrum is encoded in the quiver diagram shown in figure 
\ref{quivergen}. The RR tadpole conditions read
\beqa
\pm 3\Tr \gamma_{\theta,3}\, - \, \Tr \gamma_{\theta,7_3}\, =\, 0 
\label{rrtadpole}
\eeqa
with the upper (lower) sign corresponding to the $\ov{\rm D3}$- (resp, 
D3-) brane case.

\section{Flux Compactifications and the Klebanov-Strassler Throat}

There is much recent interest in the study of compactifications of type
IIB on Calabi-Yau (CY) threefolds (or generalisations in terms of F-theory on
CY fourfolds) with non-trivial backgrounds for NSNS and RR
3-form field strength fluxes, $H_3$ and $F_3$ respectively 
\cite{fluxes,gvw,gkp}.

These compactifications have several interesting features. One of them is
that they naturally have very few moduli. Namely, the fluxes induce a
potential for the scalar moduli of the underlying Calabi-Yau 
compactification, lifting the corresponding flat directions and 
stabilising most of them, in particular the dilaton and all complex 
structure moduli. The stabilisation can be intuitively understood as 
follows. The equations of motion require the flux combination $G_3=F_3-iS 
H_3$ (where $S=1/g_s-ia$ is the complex IIB dilaton) to be imaginary 
self-dual with respect to the underlying CY metric, $G_3=i*_{6d} G_3$. 
Since the fluxes $F_3$, $H_3$ are quantised, the flux density $G_3$ 
depends on complex structure moduli (which control the size of 3-cycles), in 
addition to the dilaton. Hence, for the imaginary self-duality condition
to hold, the dilaton and complex structure moduli must take particular
vacuum expectation values, hence they are stabilised by the the choice of
flux quanta.
 
In 4d effective supergravity terms, the fluxes generate a
superpotential \cite{gvw}:
\beq
W\ =\ \int_M G_3\wedge \Omega\,,
\eeq
where $\Omega$ is the holomorphic 3-form in the Calabi-Yau space. This 
depends implicitly on the dilaton and complex structure moduli. 
Minimisation of the scalar potential leads to the imaginary self-duality 
condition, and to the moduli stabilisation described above.                 

Since the superpotential does not depend on K\"ahler moduli, they do not
appear in the scalar potential, and they are not stabilised by the flux
background. For instance, centring on the overall K\"ahler size of the
Calabi-Yau $T$, the K\"ahler potential is
\beq
K\ =\ \tilde K (\varphi_i, \varphi_i^*) -3 \log \left( T+ T^*\right)\,,
\eeq
with $\tilde K$ the K\"ahler potential for all the other fields
$\varphi_i$ except for $T$. The supersymmetric scalar potential takes the
form
\beq
V_{SUSY}\ =\ e^K\left(K^{i\bar j} D_iW \ov{D_{j} W}\right)\,,
\eeq
with $K^{i\bar j}$ the inverse of the K\"ahler metric $K_{i \bar j}=
\partial_i \partial_{\bar j} K$ and $D_iW =\partial_i W + W\partial_i
K$ the K\"ahler covariant derivative. The contribution of $T$ to the
scalar potential through the K\"ahler potential precisely cancels the
term $-3e^K|W|^2$ of the standard supergravity potential, as usual in
no-scale models \cite{noscale}. The potential is positive definite, so
the minimum lies at $V=0$, with all the fields except for $T$ fixed from
the conditions $D_iW=0$. This minimum is supersymmetric if $D_TW=W=0$ and
not supersymmetric otherwise.
 
Additional ingredients have been proposed to stabilise the additional
K\"ahler moduli. For the case of a single modulus, it was argued in
\cite{kklt} that non-perturbative effects on a D7-brane gauge sector can
stabilise it, leading to anti-de Sitter vacua. Further addition of
anti-D3-branes could then be employed to turn the vacuum into a de Sitter
vacuum.                                     

A second important property of these compactifications is that they
naturally lead to warped metrics. Namely, the internal background metric 
is not the Ricci-flat one for the Calabi-Yau, but conformal to it, due to a
non-trivial warp factor,
\beqa
ds^2 \, =\, Z^{-1/2}\, \eta_{\mu\nu}\, dx^\mu\, dx^\nu \, + Z^{1/2}
ds^2{}_{CY}
\eeqa
The warp factor is due to the flux backreaction on the metric
\beqa
\nabla^2 Z \, \simeq \, G_{lmn}^*\, G^{lmn}
\eeqa
(where $\nabla$ and the raising of indices are done with the underlying 
CY metric $ds^2{}_{CY}$). The warp factor implies that the 4d scales of 
physical processes suffer a redshift which depends on the point of the 
internal space at which they take place. This effect can be quite strong 
in highly warped configurations, and provides the string theory realisation 
of the Randall-Sundrum hierarchy. In this reference, exponential warp 
factors were proposed in \cite{rs} as a way to solve the hierarchy between 
the weak and the Planck scale. The particular setup employs a 5d geometry, 
given by a slice of AdS$_5$ space between two $\IM_4$ boundaries, denoted 
infrared/ultraviolet regions according to their large/small redshift 
factor, respectively. The Standard Model fields were assumed to be 
localised at the infrared region, so that any Planck scale effect is 
lowered to the TeV scale due to the warp factor, explaining the hierarchy 
between the electroweak and 4d Planck scales. 

There is a simple geometry with fluxes, which illustrates both these
aspects of flux compactifications. It is the Klebanov-Strassler (KS) 
throat \cite{ks}. The underlying Calabi-Yau is the deformed conifold, a 
non-compact manifold described as the complex hypersurface
\beqa
x_1^2 \, +\, x_2^2 \,+\, x_3^2\, +\, x_4^2\, =\, \epsilon
\label{coni1}
\eeqa
in $\IC^4$. There is a complex structure modulus, whose vev is $\epsilon$.
It controls the size of the unique compact non-trivial 3-cycle in the 
geometry, which is the $\IS^3$ obtained (e.g. for $\epsilon$ real) by 
taking real $x_i$ in (\ref{coni1}). Its dual 3-cycle is non-compact, and
it is convenient to introduce a cutoff $\Lambda$ to render its volume 
finite \footnote{If the conifold is embedded in a global compactification, 
the volume of this cycle is indeed finite.}. Introducing $M$ units of flux 
$F_3$ over this $\IS^3$, and $-K$ units of $H_3$ flux over the dual 
cycle, leads to stabilisation of the complex structure modulus at the 
value
\beqa
\epsilon = \exp(-2\pi K/M g_s)
\eeqa
The full metric is described in \cite{ks}. The geometry has the structure
of a highly warped throat, looking like AdS$_5$ space (times an internal 
compact geometry $T^{1,1}=\IS^2\times \IS^3$) in its central region. 
However, the finite size of the $\IS^3$ caps off the throat at a finite distance, leading to a maximum warp factor at the bottom of the throat given 
by
\beq
Z \sim e^{8\pi K/3Mg_s}
\label{hier}
\eeq
In compactifications, the other end of the throat (the mouth) ends in the
(slightly warped, almost flat) geometry of the Calabi-Yau. The situation
is shown in figure \ref{throatCY}.

As emphasised in \cite{verlinde,gkp}, the situation is highly reminiscent 
of \cite{rs}, with a region looking like an slice of AdS$_5$ space, ending
at the infrared (by the finite size $\IS^3$) and at the ultraviolet (by
the remaining piece of the compact Calabi-Yau). In fact, this was advocated 
in \cite{gkp}, as a string theory realisation of the Randall-Sundrum 
proposal. However, a concrete realisation where there is a realistic gauge 
sector localised at the bottom of the throat has not been achieved in the 
literature. Our purpose in the present paper is to provide explicit 
examples of this kind.

\section{Orbifolds within Conifolds}

\subsection{More on the conifold}

We are interested in constructing a strongly warped throat, similar to 
the KS model, at the bottom of which we have an orbifold singularity
(preferably $\IC^3/\IZ_3$), on which we would like to place anti-D3-branes 
to lead to a chiral gauge sector. This can be achieved by starting with a Calabi-Yau geometry including a set of 3-cycles (preferably 3-spheres) and 
a $\IC^3/\IZ_3$ singularity lying on them. Introduction of large fluxes on 
those 3-cycles (and their duals) would generate a highly warped throat, at 
the tip of which the singularity sits.

A simple possibility would be to construct the quotient of the deformed 
conifold by a $\IZ_3$ action with isolated fixed points. Unfortunately, 
the deformed conifold does not admit such symmetries. For instance, we can
change variables and write (\ref{coni1}) as
\beqa
xy-uv=\epsilon
\label{con2}
\eeqa
There is a $\IZ_3$ symmetry $(x, u) \to e^{2\pi i/3} (x,u)$ and $(y,v)\to 
e^{-2\pi i/3}(y,v)$, which unfortunately is freely acting. Other possible 
$\IZ_3$ symmetries, like $x\to e^{2\pi i/3} x$, $y\to e^{-2\pi i/3} y$, 
leaving $u,v$ invariant, have a whole complex curve (defined by 
$uv=\epsilon$) of fixed points, so that locally the singularity is 
$\IC\times \IC^2/\IZ_3$. This singularity leads to $\NN=2$ world-volume 
theories on $\ov{\rm{D3}}$-brane probes, which are non-chiral and thus 
not interesting.

\medskip

Hence, it is not possible to consider a deformed conifold invariant under 
a $\IZ_3$ symmetry of the required kind. An alternative is to construct 
geometries including several deformed conifolds, arranged in an invariant 
way under a suitable $\IZ_3$ symmetry, by which we subsequently quotient 
the geometry. This is accomplished in next subsections, but before we need 
some additional information on the conifold.

\begin{figure}
\begin{center}
\centering
\epsfysize=4.5cm
\leavevmode
\epsfbox{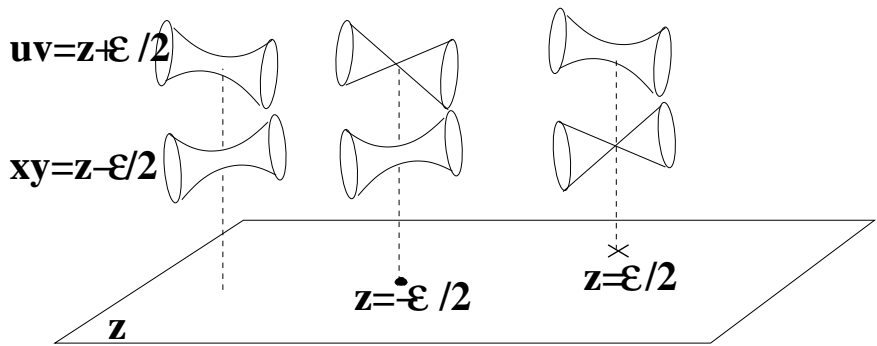}
\end{center}
\caption[]{\small Description of the conifold geometry as a double
$\IC^*$ fibration.}
\label{conifold}
\end{figure}

The deformed conifold (\ref{con2}) can be equivalently described by
\beqa
xy=z-\epsilon/2 \quad ; \quad uv=z+\epsilon/2
\eeqa
In this description, the coordinate $z$ parametrises a complex plane, at 
each point of which the coordinates $x,y$ (subject to the equation) 
describe a $\IC^*$ fibration (with a fiber topologically $\IR\times \IS^1$).
The fiber degenerates to two complex planes at the point $z=\epsilon/2$, 
where we have $xy=0$. Finally there is an analogous $\IC^*$ fibration 
parametrised by $u,v$, and degenerating at $z=-\epsilon/2$. See figure 
\ref{conifold}.
 
\begin{figure}
\begin{center}
\centering
\epsfysize=4cm
\leavevmode
\epsfbox{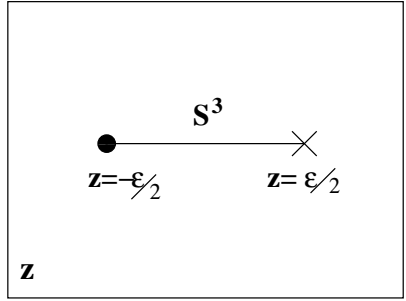}
\end{center}
\caption[]{\small Simplified pictorial depiction of the conifold and its
$\IS^3$. The dot and the cross denote the degeneration points of the two
$\IC^*$ fibrations, respectively.}
\label{sphere}
\end{figure}

The $\IS^3$ in the deformed conifold is visible as follows. In each 
of the two $\IC^*$ fibers there is an $\IS^1$, which shrinks to zero size 
at $z=\pm \epsilon/2$, respectively. Consider the 3-cycle obtained by 
taking a real segment in the $z$-plane joining the points $z=\epsilon/2$ 
and $z=-\epsilon/2$, and fibering over it the $\IS^1$ fiber in the first 
$\IC^*$ fibration and the $\IS^1$ fiber in the second $\IC^*$ fibration. 
This defines a 3-cycles without boundary, which topologically is $\IS^3$, 
and whose size is controlled by the parameter $\epsilon$. Our pictorial 
depiction of those cycles is shown in figure \ref{sphere}.
 
\subsection{Elliptic fibrations and Schoen's Calabi-Yau}

In this section we describe geometries containing several deformed 
conifolds.

\begin{figure}
\begin{center}
\centering
\epsfysize=3.5cm
\leavevmode
\epsfbox{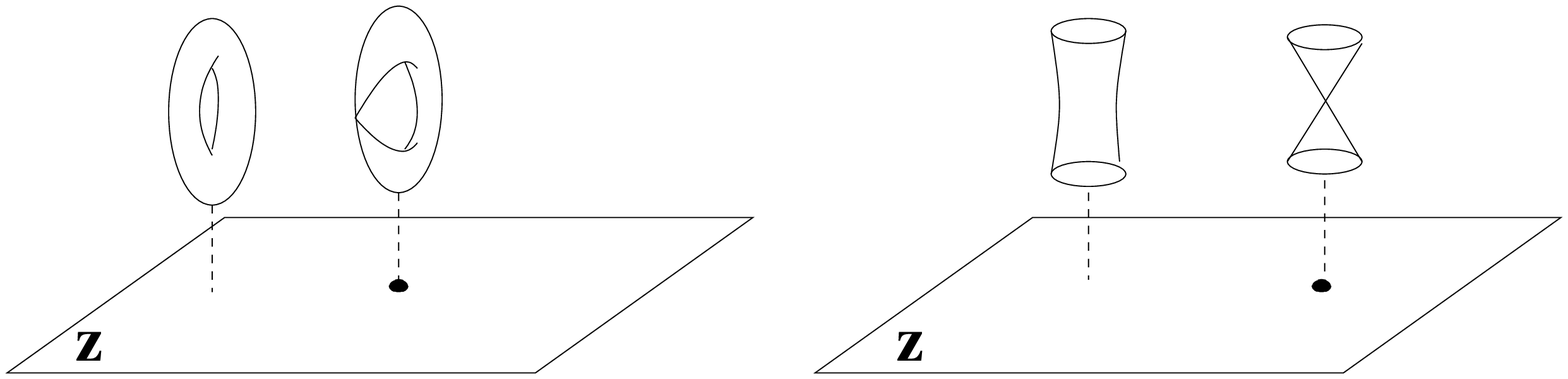}
\end{center}
\caption[]{\small The degeneration of an elliptic fibration near a
degenerate fiber can be locally described as a $\IC^*$ fibration.}
\label{pinching}
\end{figure}

Let us start by considering the two-fold defined by
\beqa
y^2=x^3 + f(z)x + g(z)
\eeqa
where $f,g$ are holomorphic polynomials (whose degree is constrained, 
although we momentarily skip this point). It describes an elliptic 
fibration over a complex plane, in Weierstrass form. Namely $z$ 
parametrises a complex plane, and at each point in $z$ the variables $x$, 
$y$ parametrise a two-torus (i.e. elliptic curve). The fiber is 
degenerate on top of the points $z$ satisfying the discriminant equation
\beqa
\Delta=4f(z)^3+27g(z)^2=0
\eeqa
at which a $(p,q)$ 1-cycle of the two-torus pinches to zero size. Near each 
of these degenerations, say at $z=z_0$, one can introduce local complex 
coordinates in the torus $u,v$, such that the geometry is locally 
$uv=z-z_0$. Namely the local geometry corresponds to a $\IC^*$ fibration 
near a degeneration point, see figure \ref{pinching}.

Hence, geometries with deformed conifold singularities can be constructed 
by considering double elliptic fibrations. Namely, consider the threefold
\beqa
y^2=x^3 + f(z)x + g(z) \quad  ; \quad
y'^2=x'^3 + f'(z)x' + g'(z)
\label{double}
\eeqa
describing two elliptic fibrations over the $z$-plane. The above equation 
can be used to define a compact Calabi-Yau manifold, by considering $z$ to 
parametrise (a patch) in $\IP_1$ (i.e. by homogenising (\ref{double}) 
with respect to projective coordinates $[z,w]$ for $\IP_1$), and imposing 
the polynomials $f$, $f'$ and $g$, $g'$ to be homogeneous of degree 4 and 
6, respectively. Such compact geometries have been studied in 
\cite{schoen}. In general, we will be interested in local models, namely 
non-compact versions of (\ref{double}), so we regard $z$ as 
parameterising a complex plane, and take $f$, $f'$, $g$, $g'$ of low 
enough degree. Schoen's manifold then provides a possible global 
embedding of these local models in a compact Calabi-Yau.

Using the above information, one can see that such geometries contain
non-trivial compact 3-cycles. Denote $z_i$, $z'_j$ the degeneration points 
of each fibration, and $(p_i,q_i)$, $(p'_j,q'_j)$ the 1-cycles in the 
corresponding elliptic fiber degenerating at those points. An $\IS^3$ is 
obtained by considering a segment in the $z$-plane joining a point $z_i$ 
with a point $z'_j$, and fibering over it the $(p_i,q_i)$ 1-cycle of the 
first elliptic fibration times the $(p'_j,q'_j)$ 1-cycle of the second. 
For geometries with coinciding $z_i$ and $z'_j$, the $\IS^3$ has zero 
size and we get a conifold singularity. If they are close but not 
coincident, then locally we have a deformed conifold, with deformation 
parameter $\epsilon_{ij}=z_i-z'_j$.
 
\subsection{The ${\bf Z}_3$-invariant manifold}
\label{finalgeom}

We would like to consider geometries invariant under a $\IZ_3$ action with 
isolated fixed points. A suitable $\IZ_3$ symmetry is provided by
$z\to e^{2\pi i/3} z$, $x\to e^{2\pi i/3} x$, $x'\to e^{2\pi i/3} x'$
Manifolds with this $\IZ_3$ symmetry are of the form
\beqa
y^2= x^3+f_2 z^2 x+ (g_2 z^6+g_1z^3+g_0) \nonumber \\
y'^2= x'^3+f_2' z^2x'+ (g_2' z^6+g_1'z^3+g_0')
\eeqa
Fixed points lie on top of $z=0$, where the elliptic fibers are $\IZ_3$ 
symmetric $\IT^2$'s, namely their complex structure parameter corresponds 
to $\tau=e^{2\pi i/3}$. The $\IZ_3$ action has three fixed points on each, 
so in total the geometry has nine isolated fixed points, around which the 
geometry is locally $\IC^3/\IZ_3$.
                           
We are interested in constructing a simple as possible geometry. Hence 
we can impose the elliptic fibrations to have just three degeneration 
points, so that the geometry, roughly speaking, contains three copies of 
the deformed conifold, rotated to each other by the $\IZ_3$. Without loss 
of generality, such geometries read
\beqa
y^2 & = & x^3 -3(z/z_0)^2 x + 2(z/z_0)^3 - 4 \nonumber \\
y'^2 & = & x'^3 -3(z/z_0')^2 x' + 2(z/z_0')^3 - 4
\label{doublefin}
\eeqa
This describes an elliptic fibration degenerating at points $z=\omega 
z_0$, and another elliptic fibration degenerating at $z=\omega z_0'$, 
where $\omega=1$, $e^{2\pi i/3}$, $e^{-2\pi i/3}$. Clearly $z_0$ or $z_0'$ 
can be eliminated by a rescaling of $z$, but we prefer to keep the above 
more symmetric description of the manifold. In the above geometry, the
1-cycles degenerating at the different point, may be chosen to be
the cycles $(2,-1)$, $(-1,2)$ and $(-1,-1)$, and analogously for the 
primed fibration. This double elliptic fibration was considered (in a 
different context) in section 3.2 of \cite{uralocal}.

\begin{figure}
\begin{center}
\centering
\epsfysize=3.5cm
\leavevmode
\epsfbox{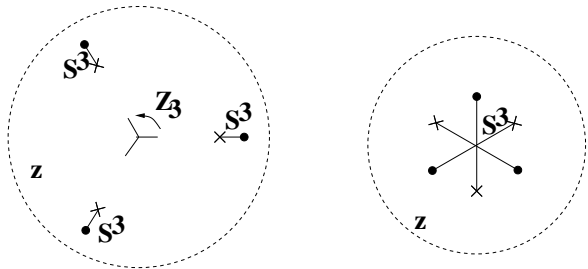}
\end{center}
\caption[]{\small The first figure shows the a general $\IZ_3$-invariant 
configuration for the three deformed conifolds. In the second figure 
we show a configuration invariant under an additional $\IZ_2$ symmetry.}
\label{final}
\end{figure}

A pictorial depiction of geometries of this kind are shown in figure 
\ref{final}. The first figure is the general kind of configuration. The 
second shows the geometry at a point in moduli space where it has an 
additional $\IZ_2$ symmetry, given by $x,y\to x',y'$, $z\to -z$, that we 
will exploit below. Our final geometry is obtained by modding out the manifold 
(\ref{doublefin}) by the $\IZ_3$ symmetry, which has nine fixed points.
 
\subsection{Introduction of D-branes}

One can now introduce D-branes in this local geometry, to obtain chiral gauge
sectors. In particular, we will introduce D3- or $\ov{\rm D3}$-branes 
sitting at the $\IC^3/\IZ_3$ orbifold points, and D7-branes located at 
$z=0$, hence passing through the singularities, and wrapped on the 
$\IT^4$ fiber in the geometry. These sets of D-branes lead to a chiral 
open string spectrum localised on their world-volume. The models are such 
that the Standard Model group and matter multiplets are located on D3- or 
$\ov{\rm D3}$-branes at the origin in the $z$-plane and elliptic fibers. 

The simplest 
option, which already provides a chiral gauge sector, is to locate a 
number $n$ of $\ov{\rm D3}$-branes at the orbifold point at the origin. It 
leads to a $U(n)^3$ world-volume gauge theory with $\NN=1$ supersymmetry 
(ignoring the effect of the fluxes) and chiral matter multiplets in the 
three copies of the representation $(n,\bar{n},1)+(1,n,\bar{n})+
(\bar{n},1,n)$. 

Although chiral, this theory is not realistic. More interesting models can 
be obtained exploiting a further natural ingredient in flux 
compactifications, namely by introducing D7-branes, passing through the 
origin. Their presence allows to consider richer Chan-Paton structures for 
the $\ov{\rm D3}$-branes, consistently with twisted RR tadpole 
cancellation, and to build models with gauge group and matter closer to 
the Standard Model. Explicit D-brane configurations of this kind are 
discussed in section \ref{realistic}.

In the next subsection we describe the introduction of fluxes in
this geometry, in order to create a highly warped KS-like throat, and 
consider a choice of fluxes which ensures (via moduli stabilisation) that 
the singularities (and hence the D-brane configuration and the gauge 
sector) sit at the bottom of the highly warped throat. 

An important question is the backreaction of the branes on the underlying 
geometry. This can be considered small as long as the flux quanta are 
larger than the number of D-branes required, a situation which can be 
achieved consistently with the desired hierarchy (e.g. by rescaling the 
fluxes keeping the hierarchy (\ref{hier}) constant). A second important 
point is that the chiral part of the D-brane world-volume spectrum can be 
read off from the configuration without fluxes, since is protected by 
chirality. Non-chiral fields massless in the absence of fluxes, may 
however receive flux-induced mass terms.

\subsection{Turning on Fluxes and the Structure of the Throat}
\label{fluxesthroat}

We would now like to turn on 3-form fluxes on the 3-cycles in our geometry, 
and build a highly warped KS-like throat. We choose to introduce RR 
fluxes on the compact 3-cycles and NSNS fluxes on non-compact 3-cycles in 
the geometry. The flux we introduce stabilises the complex structure 
modulus (that can be considered to be $z_0/z_0'$), and leads to a geometry 
with small $\IS^3$s, lying at the bottom of the throat. 

Moreover, we would like the final geometry to contain the $\IC^3/\IZ_3$ 
orbifold singularity at the bottom of the throat. This is ensured if the 
final value of $z_0/z_0'$ is such than the origin is on some $\IS^3$ with 
non-zero flux. Figure \ref{final}a shows a situation where this would not be 
the case, namely where the throat develops and the $\IC^3/\IZ_3$ singularity 
is left behind, in the almost unwarped part of the Calabi-Yau. On the 
other hand, figure \ref{final}b shows a situation where there are 
3-spheres passing through the origin. This implies that, when we introduce 
fluxes on them and the throat develops, the origin will lie at the bottom 
of the throat.

In general, and for such a complicated geometry, it is difficult to
determine the value at which moduli stabilise, in terms of the flux
quanta. In this paper we do not solve this problem systematically, but
rather notice that the interesting case in figure \ref{final}b is
invariant under the $\IZ_2$ symmetry discussed at the end of section
\ref{finalgeom}. It is then expected that a $\IZ_2$-invariant set of flux 
quanta assignments will ensure the moduli stabilise at precisely such 
geometry.

\begin{figure}
\begin{center}
\centering
\epsfysize=6.5cm
\leavevmode
\epsfbox{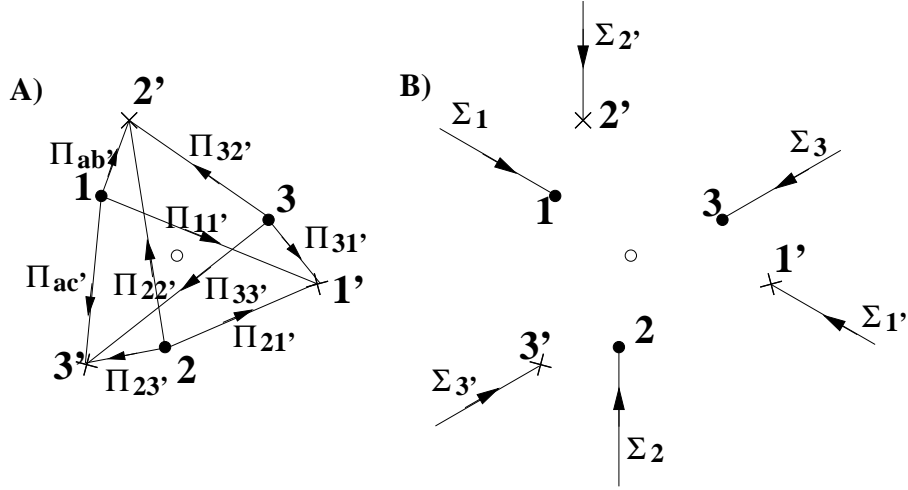}
\end{center}
\caption[]{\small Pictorial depiction of some interesting 3-cycles in our 
geometry. Figure 7a shows the compact 3-cycles, in three different orbits 
of the $\IZ_3$ action. Figure 7b shows the non-compact 3-cycles, which 
lie in two $\IZ_3$ orbits.}
\label{star}
\end{figure}

Since we are not interested in the most general flux, we do not require
a full knowledge of a basis of 3-cycles in the geometry. The only requirement 
in introducing the fluxes is the consistency condition that homologically 
related 3-cycles get similarly related fluxes. In our geometry there are 
three kinds of $\IS^3$'s shown in figure \ref{star}a, corresponding to 
three $\IZ_3$ orbits. We denote by $\Pi_{ij'}$ ($i,j=1,2,3$) the 3-cycle 
defined by the segment joining the degenerations with labels $i$ and 
$j'$. The three classes are $\Pi_{11'}$ (and its images $\Pi_{22'}$, 
$\Pi_{33'}$), $\Pi_{12'}$ (and images $\Pi_{23'}$, $\Pi_{31'}$) and 
$\Pi_{13'}$ (and images $\Pi_{21'}$, $\Pi_{32'}$). There are also some 
non-compact 3-cycles, shown in figure \ref{star}b, which 
lie in two $\IZ_3$ orbits. They are denoted $\Sigma_i$ and $\Sigma_{i'}$\footnote{Of course, in order to fully determine a 3-cycle we should not only specify the segment in the base, but also the two 1-cycles in the two elliptic fibrations. The 3-cycles that we call $\Sigma_i$ and $\Sigma_{i'}$ (which enter e.g. in expression \ref{homrelt}) have very precise 1-cycles fibered, which are discussed in detail in appendix A.}.

The $\IZ_3$ acts as a simultaneous cyclic rotation of $(1,2,3)$ and 
$(1',2',3')$, and leaves the geometry invariant. The $\IZ_2$ action is
$[\Pi_{ij'}]\to [\Pi_{ji'}]$, $[\Sigma_i]\leftrightarrow [\Sigma_{i'}]$, 
and is a symmetry of the homology lattice, but in general not a symmetry 
of the geometry. For $z_0=-z_0'$, the $\IZ_2$ is a symmetry of the 
geometry as well.

As suggested above, there are several homology constraints among the above 
set of 3-cycles, which our flux assignments should respect. Some of the 
relevant ones for us are described in appendix A, in terms of deformation arguments. The basic relation reads
\beqa
3[\Sigma_{3}]-3[\Sigma_2]=[\Pi_{11'}]+[\Pi_{12'}]+[\Pi_{13'}]
\label{homrelt}
\eeqa
and similar ones among their $\IZ_3$ images.
This implies that if the flux over $[\Sigma_2]$ and $[\Sigma_3]$ are  
equal (as required by $\IZ_3$ symmetry), then the sum of fluxes over 
$[\Pi_{11'}]$, $[\Pi_{12'}]$, $[\Pi_{13'}]$, must vanish. If we are 
interested in a $\IZ_2$ invariant assignment of fluxes, then (using also 
the $\IZ_3$ symmetry) the flux over $[\Pi_{12'}]$ and $[\Pi_{13'}]$ must 
be equal, and then the flux over $[\Pi_{11'}]$ must be (minus) twice that 
amount. Similar statements follow for the $\IZ_3$-related images of these 
cycles.

\begin{figure}
\begin{center}
\centering
\epsfysize=5.5cm
\leavevmode
\epsfbox{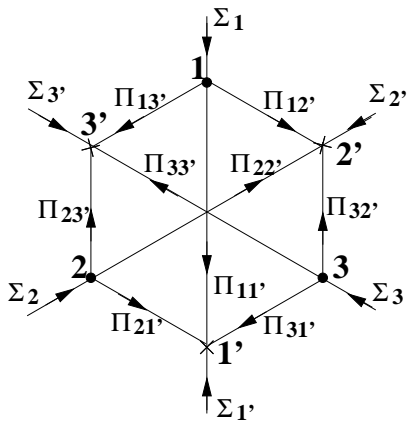}
\end{center}
\caption[]{\small The figures shows the set of 3-cycles on which we have 
turned on a $\IZ_2$-invariant set of 3-form fluxes. The moduli stabilise 
at the shown $\IZ_2$-invariant geometry.}
\label{z2inv}
\end{figure}

The final configuration of fluxes is therefore
\beqa
& \int_{[\Pi_{ii'}]}\, F_3 = K \quad , \quad
\int_{[\Pi_{ij'}]}\, F_3 =-K/2 \, \quad \quad {\rm for }\,\, i\neq j & 
\nonumber \\
& \int_{[\Sigma_{i}]} H_3 =  -\int_{[\Sigma_{i'}]} H_3 = M &
\eeqa
and succeeds in providing a set of $\IZ_2$-invariant fluxes. Hence they 
stabilise the moduli at the $\IZ_2$-invariant geometry, namely $z_0=-z_0'$.
In such situation, the orbifold points in the $\IZ_3$ quotient lie on the 
3-sphere  associated to the orbit $\Pi_{ii'}$, so that the orbifold point 
is at the bottom of the throat. The final picture for the 3-cycles with 
flux, and the resulting geometry, is in figure \ref{z2inv}.

\begin{figure}
\begin{center}
\centering
\epsfysize=4.5cm
\leavevmode
\epsfbox{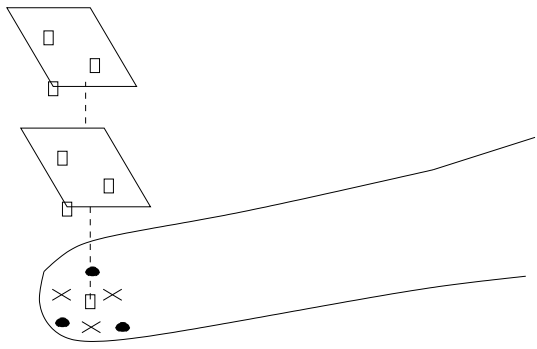}
\end{center}
\caption[]{\small Structure of the throat in the $z$-plane, over which we 
have a double elliptic fibration. At its tip, round dots and crosses 
denote degenerations of the elliptic fibers, while square dots represent 
fixed points of the $\IZ_3$ action.}
\label{tipstructure}
\end{figure}

It is also important to notice that the different $\IS^3$ (related by 
$\IZ_3$) in the covering space wrap in different directions in the 
$\IT^4$, and that their size is thus related to the local size of $\IT^2$
fibers. Hence it is reasonable to expect the full $\IT^4$ in the fiber 
to sit at the bottom of the throat. Schematically, the $z$-plane 
elongates producing the throat, all over which we have a $\IT^4$ 
fibration. At the bottom of the throat, the double elliptic fibration has 
several degenerations, giving the $\IZ_3$ related $\IS^3$'s. The 
non-compact direction of the $z$-plane plays the role of the radial 
direction in the KS-like throat, so that there is large variation of the 
warp factor along that direction, and the central piece of the throat is 
an slice of AdS$_5$. In contrast with the KS throat, the internal space 
along the throat is not $T^{1,1}$, but rather a $\IT^2\times \IT^2$ 
bundle over $\IS^1$. A picture of the throat and the structure at its tip 
is given in figure \ref{tipstructure}. 

We have thus succeeded in constructing a geometry and flux background 
leading to a highly warped throat with orbifold singularities at its 
bottom. The geometry is rather involved, hence we cannot provide the 
explicit metric and flux density profile. We however expect that, given 
the analogy, its structure is similar to the KS solution. Further work to 
quantify this would clearly be desirable.
                 
\section{Examples of Realistic Models}
\label{realistic}
 
We can now start describing some of the realistic models we can build. One 
can greatly benefit from the analysis in \cite{aiqu}, which gives us the model 
building rules based on a local analysis, and can be applied to the throat 
we have just constructed.

\subsection{Two Scenarios}

As mentioned in the introduction we may consider two main scenarios
\footnote{A general discussion of the different possible scales
  appearing in these classes of models can be found in \cite{gw}.}

\begin{enumerate}

\item{}
The Standard Model is embedded on a stack of anti D3-branes and D7-branes. 
In this construction, supersymmetry is explicitly broken on the tip of the 
throat. There 
are different versions of this scenario (as illustrated in some of the models)
where the other branes on the throat are D3- and D7- and/or $\ov{\rm D3}$-,
$\ov{\rm D7}$-branes, but their main properties are similar. The 
fundamental scale at the location of these Standard Model branes should 
then be of the order of $1$ TeV. This can be easily achieved from the 
strong redshift factor at the bottom of the throat, as in the 
Randall-Sundrum case. In this construction, most of the phenomenological 
analysis carried out for Randall-Sundrum models applies \cite{rubakov}. 
This scenario has the advantage that the anti D3-branes are dynamically 
attracted by the fluxes and naturally fall to the bottom of KS-like
throats \cite{kpv}. This reduces the number of moduli, but more 
importantly provides an explanation for the location of the Standard 
Model fields at the strongly redshifted region.

\item{}
It is possible to construct supersymmetric models in the present setup, 
namely, if the standard model is constructed inside a stack of D3-branes. 
Both the D7- and D3-branes preserve the same supersymmetry and we have a 
version of the MSSM at the end of the throat. Additional D-branes at 
orbifold points away from the origin may be D3- or $\ov{\rm D3}$ branes. 
In the first case supersymmetry is fully preserved on the throat, if the 
fluxes also preserve it. The background outside the throat may preserve 
supersymmetry as well (e.g. as in the F-theory example in section 4.3 in 
\cite{aiqu}) or break supersymmetry (as in most models in \cite{aiqu}) by 
e.g. distant antibranes, non-supersymmetric fluxes in the CY away from 
the throat, etc. In both of these cases, supersymmetry breaking 
would be implemented via gravity mediation from a hidden sector, given by 
the distant non-supersymmetric source in the latter case, and from e.g. 
gaugino condensation or other non-perturbative effect in the former. The 
amount of warping on the throat in this case is not related to the 
hierarchy (which is stabilised by supersymmetry) and its choice would 
depend on the supersymmetry breaking mechanism.

\end{enumerate}

\subsection{Left-Right Models}

\subsubsection{Models with anti D3-branes}

Consider a set of $\ov{\rm D3}$-branes at the origin with
\beqa 
\gamma_{\theta,\bar{3}}= \diag(\id_3, \alpha \id_2,\alpha^2 \id_2)
\eeqa
In order to cancel RR twisted tadpoles, we add D7-branes wrapped 
on the $\IT^4$ fiber and passing through the $\IZ_3$ point, with
$\gamma_{\theta,7}=\id_3$. This cancels the RR twisted tadpole at the origin (\ref{rrtadpole}), but introduces tadpoles at the other fixed points. They can be cancelled by locating one anti-D3-brane at each, with $\gamma_{\theta,\bar{3}_i}= \id_1$.

This leads to a consistent model, which has a LR sector at the end of the 
throat. The spectrum in the sector of antibranes at the origin is 
essentially $SU(3)\times SU(2)_L\times SU(2)_R\times U(1)_{B-L}$. The 
additional $\ov{\rm D3}$-branes give some abelian factors, which disappear as 
discussed below. The D7-branes in principle lead to an $U(3)$ group, but 
the $U(1)$ factor is anomalous and becomes massive, while the $SU(3)$ factor 
provides a (gauged) horizontal symmetry for leptons. The full 
spectrum is shown is table \ref{tabpslr} of appendix B.

There remains the crucial issue of $U(1)$ anomalies.
As discussed in \cite{aiqu}, a `diagonal' combination of the $U(1)$ 
factors arising from the D3-branes at the origin is non-anomalous and 
remains massless. It is given by
\beqa
Q_{B{\rm -}L}= - 2 \left(\frac 13\ Q_{3}\ +\frac 12  \ Q_{L}\ +
\frac12  \ Q_{R}\right)
\label{bminusl}
\eeqa
and provides the familiar $B-L$ quantum number for fields in the LR 
sector.

The remaining two $U(1)$'s and the $U(1)$ from the D7-branes are anomalous 
(with anomaly cancelled by the Green-Schwarz mechanism in \cite{gsorbifold}), 
and become massive. Finally, the additional $U(1)$ factors from the extra 
$\ov{\rm{D3}}$-branes have no charged fermions, and are automatically 
anomaly-free. However, they become massive as well, since they have non-zero 
$B\wedge F$ couplings with twisted closed string moduli (in analogy with 
the discussion in \cite{imr}). In the mentioned table we provide the charges 
under all $U(1)$ factors, but list in the last column the charges under 
the only surviving factor (\ref{bminusl}).

Since we have anti D3's, the model is non-supersymmetric, as is manifest in 
the spectrum. However, supersymmetry is not essential in the phenomenology 
of the models, since the warped throat stabilises the hierarchy, a la 
Randall-Sundrum. This is the first semi-realistic model where the RS 
mechanism is put to work in an explicit string theory context.

Notice that the global geometry is different from the toroidal models in 
\cite{aiqu}. This has implications (which however do not affect the chiral 
structure of the LR model just constructed). For instance, it is not 
possible to turn on Wilson lines on the D7-brane to further break the 
$SU(3)$ horizontal symmetry. In our geometry, the 1-cycles in the $\IT^4$ 
fiber, on which the D7-branes wrap, are contractible so there is no 
moduli associated to a choice of Wilson lines. An alternative to break the 
D7-brane gauge group would be to introduce D7-brane world-volume magnetic 
fields, which are quantised and thus discrete; however we do not discuss 
this possibility in the present paper.

\subsubsection{Models with D3-branes}

Consider instead a set of D3-branes at the origin with
\beqa 
\gamma_{\theta,3}= \diag(\id_3, \alpha \id_2,\alpha^2 \id_2)
\eeqa
Again, in order to cancel RR twisted tadpoles, we add D7-branes with
$\gamma_{\theta,7}=\diag(\alpha \id_3, \alpha^2 \id_3)$, and one D3-brane 
at each of the other fixed points, with $\gamma_{\theta,3_i}= \id_1$
This leads to a consistent supersymmetric model, which has a LR sector at the end of the throat.

\subsection{Standard Model Examples}

It is also possible to construct local configurations with the Standard Model 
gauge group and 3 families of chiral fermions. The essential features of 
this class of models have been discussed in \cite{aiqu}. In order to 
illustrate their basic structure, let us consider the following particular 
example.

\subsubsection{The Standard Model on anti D3-branes}

Consider placing $\ov{\rm D3}$-branes at the origin, with the Chan-Paton 
embedding
\beqa \label{smeq}
\gamma_{\theta,\bar{3}}= \diag(\id_3, \alpha \id_2,\alpha^2 \id_1)
\eeqa  
The twisted tadpole is cancelled at the origin by adding D7-branes, 
wrapping the fiber $\IT^4$ and with a Chan-Paton embedding
\beqa 
\gamma_{\theta,7}= \diag(\id_6, \alpha \id_3)
\eeqa
Again, these generate a twisted tadpole at the other fixed points, that 
we compensate by the addition of $\ov{\rm D3}$-branes with
\beqa 
\gamma_{\theta,\bar{3}_i}= \diag(\id_2, \alpha \id_1)
\eeqa
The above configuration leads to the Standard Model gauge group 
$SU(3)\times  SU(2)_{L}\times U(1)_{Y}$ living in the world-volume of the 
$\ov{\rm D3}$-branes at the origin. Several of the extra $U(1)$ factors of 
the anti-branes and from the D7-branes become massive as discussed above. 
The linear combination given by
\beqa
Q_Y=- \left (\frac 13 Q_3 \, + \, \frac 12 Q_2 \, +\, Q_1 \right )
\eeqa
is anomaly-free, remains massless, and exactly reproduces the SM 
hypercharge.

The $SU(6)\otimes SU(3)$ gauge group coming from the D7's represents an 
extra gauged symmetry. As pointed out in \cite{aiqu}, the particular 
structure of this kind of models allows to potentially break that hidden 
gauge symmetry (by giving the appropriate vevs to scalar fields in the 77 
sector) without spoiling the good hypercharge for the matter fields. This 
breaking may also give a mass to some exotic matter in the 37 sector. In 
any case, these more detailed properties are model dependent and we skip 
them in our discussion.

The complete spectrum for this model is displayed in table \ref{tabpsSM} 
of the appendix.

Clearly, the model we have worked out is just an example. There exist 
several other possibilities, like cancelling twisted tadpoles with 
$\ov{\rm D7}$-branes or adding D3-branes instead of $\ov{\rm D3}$-branes 
outside the origin. In any event, since the local structure around the 
origin is close to the above, most models are similar and inherit the 
same appealing properties.

\subsubsection{The Standard Model on D3-branes}
 
Let us now describe a similar construction using D3-branes instead. 
We introduce a set of D3-branes at the origin, with
\beqa
\gamma_{\theta,3}= \diag(\id_3, \alpha \id_2,\alpha^2 \id_1)
\eeqa  
The twisted tadpole is cancelled at the origin by adding D7-branes, 
with
\beqa 
\gamma_{\theta,7}= \diag(\alpha \id_3,\alpha^2 \id_6)
\eeqa
and D3-branes away from the origin with
\beqa 
\gamma_{\theta,{3}_i}= \diag(\id_2, \alpha \id_1)
\eeqa
By arguments similar to the above one, this
 configuration leads to the Standard Model gauge group $SU(3)\times  SU(2)_{L}\times U(1)_{Y}$ living in the worldvolume of the D3-branes at the origin. Its spectrum is displayed in table \ref{tabpssm1}.

\subsection{Pati-Salam Examples}

It is also interesting to consider other model building possibilities. As 
an illustration, we construct a Pati-Salam model. Consider a set of anti 
D3-branes at the origin, with  
\beqa 
\gamma_{\theta,\bar{3}}= \diag(\id_4, \alpha \id_2,\alpha^2 \id_2)
\eeqa
To cancel RR twisted tadpoles, we introduce D7-branes with
$\gamma_{\theta,7}= \diag(\id_6)$, and anti D3-branes at other fixed points, with $\gamma_{\theta,\bar 3_i}= \diag(\id_2)$.

The spectrum in the sector of the antibranes at the origin is essentially 
$SU(3)\times SU(2)\times SU(2)\times U(1)_Q$. The non-anomalous $U(1)$ in 
this case corresponds to $Q=\frac 14 Q_4+\frac 12 Q_L + \frac 12 Q_R$, and 
has no natural interpretation in terms of Pati-Salam grand unification.
The full spectrum is in table \ref{tabps3r} of the appendix.
 
\section{K\"ahler Moduli Stabilisation}

\subsection{D-branes vs anti D-branes}

In the previous section we introduced two main scenarios, distinguished 
by whether the Standard Model is embedded on a stack of D3-branes or 
$\ov{\rm D3}$-branes. In both scenarios we have several moduli, which are 
not stabilised by the fluxes. For instance, both contain geometric K\"ahler 
moduli, which include the overall CY volume (plus possibly others  
controlling the size of the elliptic fibers, whose existence depends on 
the global CY structure), and two blow-up moduli at each of the orbifold 
points of the geometry. These will be discussed in next subsection. In 
addition, in scenario 1, the Standard Model branes have three complex 
moduli, associated to the possibility of combining three fractional 
D3-branes into a dynamical regular D3-brane, and moving it off the 
orbifold point. This corresponds to a flat direction of the D3-brane 
gauge theory (breaking the gauge group to an unrealistic $U(2)\times 
U(1)$), which is not lifted by the flux background, since it does not 
generate soft terms on D3-branes.

In this sense scenario 2 is more attractive, because fluxes do stabilise 
the anti D3-branes at the bottom of the throat, i.e. they generate soft 
masses and lift the latter flat direction; therefore in this case the 
standard model group is stable. Notice that this is already a concrete 
improvement over the models in \cite{aiqu}. 

Finally, it is worth mentioning D7-brane moduli. Since only fractional 
D7-branes are involved in our constructions, moduli associated to D7-brane
positions are projected out by the orbifold. Even if they were present, it 
is expected \cite{tt,ciu} that they acquire flux-induced masses, even for 
imaginary self-dual flux backgrounds.

\subsection{K\"ahler Moduli Stabilisation}
\label{modustab}

Introduction of fluxes in general stabilises the dilaton and complex 
structure moduli, as we have discussed above. On the 
other hand, K\"ahler moduli are not stabilised, and remain unfixed (at leading 
order in $\alpha'$). In order to achieve a fully realistic model, it is 
interesting to propose and device mechanisms to stabilise K\"ahler moduli. 
For simple  situations, where the only K\"ahler modulus is the overall CY 
volume, it has been proposed in \cite{kklt} that non-perturbative 
effects (like gaugino condensation in a 7-brane sector) may lead to its 
stabilisation. Moreover, full moduli stabilisation is interesting, since 
it is a requirement to achieve the construction of de Sitter vacua, which 
may be relevant to describe cosmologically interesting scenario.

In our models, the number of K\"ahler moduli is relatively large. There 
are two blow-up modes at each $\IC^3/\IZ_3$ singularity, 
adding up to a total of eighteen. In addition, when embedded in a compact 
space, there may be additional K\"ahler moduli associated to the size of 
the base and the sizes of the elliptic fibers. 

In principle, it is possible to imagine that the latter are stabilised 
by a combination of gaugino condensates associated to D7-branes away 
from the throat as in references \cite{kklt,egq,bkq}.
 This is then doable but it is clearly very model dependent, and we will not 
discuss it any further. On the other hand, the K\"ahler moduli associated 
to blow-up modes should be stabilised by a mechanism present at the 
bottom of the throat. In principle, these K\"ahler moduli contribute to 
the gauge coupling of fractional branes sitting at the singularity. For 
instance, for D3-branes associated to the eigenvalue $e^{2\pi i\, ki/N}$ 
in $\gamma_{\theta^k}$ (denoted $i^{th}$ fractional D-branes in what 
follows), the gauge kinetic function is given by 
\beqa
f(S,M_k)=\frac{1}{N}\, \left(S+\sum_{k=1}^{N-1} \, e^{2\pi i\, ki/N} \, M_k\right)
\eeqa
where $M_k$ is the K\"ahler modulus in the $k^{th}$ twisted sector. Hence, 
if the world-volume gauge theory develops a gaugino condensate, it may 
lead to the (at least partial) stabilisation of these moduli.

However, it seems that a large number of such gaugino condensates is 
required, and moreover the gauge groups on branes at singularities in our 
explicit models do not necessarily have such kind of non-perturbative effects.

There is however an alternative mechanism of stabilisation of twisted moduli,
which is automatically implemented in a large class of models, including ours.
For concreteness, let us centre on a set of $\ov{\rm D3}$-branes at an 
orbifold singularity. Following \cite{dm}, consider a complexified Kahler moduli $M_k$ in the $k^{th}$ twisted sector, and define the linear combinations
\beqa
\Phi_i=\sum_k e^{2\pi i ki/N} M_k
\eeqa 
then there is a coupling between the 2-forms $B_i$ which are the 4d duals of the RR piece in $\Phi_i$, and the $U(1)$ gauge boson on the $i^{th}$ fractional $\ov{\rm D3}$-brane, given by
\beqa
\int_{\ov{\rm D3}} \, B_i \wedge F_i
\eeqa
These couplings play an essential role in the Green-Schwarz cancellation 
of mixed $U(1)$ - non-abelian gauge anomalies \cite{gsorbifold}. Due to 
the $\NN=1$ supersymmetry unbroken by the $\ov{\rm D3}$-brane and the 
orbifold, there is a non-trivial Fayet-Illiopoulos term, given in terms 
of the scalars $\xi_i$, which are the NSNS piece of $\Phi_i$
\beqa
\int_{\ov{\rm D3}} \, \xi_i D_i
\eeqa
This leads to a D-term potential
\beqa
V_D\ = \ \sum_i\, \left(\, \sum_a\, q_{I,a}\, \Psi_{i,a}\, \Psi_{i,a}^* \, +\, \xi_i\,\right)^2
\eeqa
where $a$ runs through all the scalars $\Psi_{i,a}$, charged (with charge
$q_{i,a}$) under the $i^{th}$ $U(1)$ gauge group.

In the absence of such scalars, the D-term potential leads to a mass term
for the twisted moduli, and freezes them at zero vev (corresponding to the
blown-down limit). On the other hand, if there are massless scalars with 
charges of sign opposite to the FI term, turning on the FI term triggers 
a vev for them, leading to a restabilisation of the vacuum, usually 
restoring supersymmetry. In this case, each contribution to the D-term 
potential leads to the stabilisation of a combination of the scalars 
$\Psi_{i,a}$ and the twisted moduli. Some combination of fields remains 
as an unstabilised direction.

One could think that this is our situation, since the $\bar{3}7$ sectors do
contain scalars with charges of both signs under the $\ov{\rm D3}$-brane
world-volume $U(1)$ gauge symmetries. Indeed, in the absence of 3-form fluxes,
these scalars are massless and a non-zero FI term triggers a vev for them,
leading to a vacuum restabilisation and restoration of supersymmetry. In the
new vacuum, the fractional $\ov{\rm D3}$-branes are diluted as (anti)instantons
on the D7-branes. However, the presence of 3-form fluxes modifies the picture
because in general they induce mass terms for these fields. This mass is given
by the flux density at the orbifold point, and prevents these fields 
from acquiring a vev, at least for small FI terms, i.e. for small vevs 
for twisted moduli. Hence, there is a D-term plus flux-induced potential 
leading to a local minimum where both twisted moduli and charged scalars 
are stabilised. The masses they typically acquire are of order the string 
scale for twisted moduli, and of order the flux scale for the charged scalars.

On general grounds, to achieve full stabilisation the model should 
contain as many $U(1)$ gauge fields as twisted moduli in the model, so 
that there are enough contributions to the D-term potential. In  
constructions like the SM in section 5.3 this seems to be the case, 
and full stabilisation is plausible. On the other hand, in the LR model in section 5.2, there are too few $U(1)$'s, and some additional 
mechanisms would be required.

\medskip

Using this new ingredient one can in principle achieve models with realistic
chiral gauge sectors, and stabilisation of twisted moduli. At energies below
the flux scale, the only moduli that remains is the overall K\"ahler moduli (and
possibly other untwisted moduli), so that one recovers the situation in
\cite{kklt}. Fixing the additional moduli via non-perturbative effects one
can in principle achieve the construction of de Sitter string vacua.

\section{Phenomenology Aspects}

Even though a full phenomenological analysis of these models is beyond
the scope of this article, we will address here  several phenomenological properties of these scenarios.

\begin{enumerate}

\item
{\it Gauge Couplings}

In the scenario with the standard model on antibranes, the scale at that 
brane is close to the TeV scale and then there is no natural way to expect 
gauge coupling unification as in the MSSM \footnote{Unification in large 
extra dimension scenarios, not necessarily related to string theory, were
discussed in the unwarped case in \cite{ddg}, which achieved fast 
unification from power-law, rather than logarithmic, running in 
the extra dimensions. In the warped case it was addressed in 
\cite{alex}, where there is logarithmic running due to the 
warp factor. However, both discussions require gauge bosons living in 
the bulk, which is not the case in our models.}. Hence the relative 
values of the gauge couplings at that scale should directly correspond to 
the experimental low-energy values. For a general $\IZ_N$ orbifold with SM 
group $U(3)\times U(2)\times U(1)^{N-2}$, hypercharge is 
given by
\beqa
Q_Y&=& - \left( \frac{1}{3} Q_3 +\frac{1}{2} Q_2 + 
Q_1 \right),
\label{qdiagsm}
\eeqa
hence its normalisation depends on the order of the twist $N$. In fact, as 
it was observed in \cite{aiqu}, by normalising $U(n)$ generators such that 
${\rm Tr} T_a^2=\frac{1}{2}$, the normalisation of the $Y$ generator is 
\begin{equation}
\label{Ynorm}
k_1= 5/3 +2(N-2)
\end{equation}
This amounts to a dependence on $N$ in the Weinberg angle, namely \cite{aiqu}:
\begin{equation}
\label{sw}
\sin^2 \theta _W\ =\ \frac{g_1^2}{g_1^2+g_2^2}\ =\ 
 \frac{1}{k_1+1}= \frac{3}{6N-4}
\end{equation}
For the interesting case for us, $N=3$, we find $\sin^2\theta_W=3/14=0.2143$, which is already remarkably close to the experimental value $(0.23113\pm 0.00015)$. Hence in these models quantum corrections are less important than in
standard SUSY GUT's, for which $\sin^2\theta_W=3/8=0.375$ at tree-level and the
quantum effects are responsible for bringing it to the experimentally observed value.

In reference \cite{aiq1,aiq2,aiqu} the intermediate scale scenario was
emphasised, since the matter spectrum and the normalisation of hypercharge 
were such that in LR models unification is achieved at that scale, with 
an accuracy comparable to the SUSY GUT's in the MSSM. In our construction, 
the intermediate scale scenario is only realisable with the Standard Model 
embedded on D3-branes, so their spectra are quite similar. A detailed 
analysis of their possible gauge unification is beyond the scope of the 
present paper, but we expect similar conclusions for such models.

\item
{\it Proton Stability}

Proton stability is an important issue given that the standard model
scale is at the TeV. In the models presented, baryon number is provided
by the $U(1)$ subgroup of the color $U(3)$, and remains an exact global 
symmetry to all orders in perturbation theory \footnote{This continuous global symmetry includes the $\IZ_2$ symmetry 
mentioned in \cite{aiqu}, but implies stronger constraints. Thus our 
statements here correct the discussion in \cite{aiqu}.}. The reason for this
is that, like in the heterotic string, in D-brane models, anomalous
(and
some 
non-anomalous) $U(1)$'s become massive from the 4D realisation of the
Green-Schwarz mechanism. But unlike the heterotic case, the FI term 
may vanish and there is no need for any charged matter field to get a
vev. The corresponding $U(1)$ survives as a global symmetry (for a
detailed discussion of this issue see for instance \cite{iq}).
 Hence, any 
violation of baryon number is non-perturbative, and is therefore 
exponentially suppressed.

\item
{\it Yukawa Couplings}

The structure of Yukawa couplings may be discussed as in \cite{aiqu}, if 
we momentarily ignore the effect of fluxes. In configurations of D3- (or 
anti D3)-branes at singularities, the Yukawa couplings are provided by a 
set of gauge invariant cubic couplings between fields in the 33 or 37 (or 
$\bar 3\bar 3$ and $\bar 3 7$) open string sectors, in two possible 
structures: $(33)_1\cdot (33)_2\cdot (33)_3$, and $(33)_3\cdot (37)\cdot 
(73)$, where the subindex denotes the associated complex plane. The first 
structure is responsible for providing masses for the up quarks (and the 
down quarks as well in LR models), and as discussed in \cite{aiqu} leads to 
two degenerate and one massless quark, as in standard untwisted sectors of 
orbifold models \cite{finq}. For models of D3-branes, the couplings of 
the second form have been described in \cite{aiqu}, and lead to down quark 
masses in SM models. An important drawback is that neither those LR nor 
SM cases allow for lepton masses, since there is no gauge invariant cubic 
term  between the down type Higgs and the leptons. In this respect, 
models of $\ov{\rm D3}$-branes provide some improvement, since they lead 
both to down quarks masses and lepton masses, due to the different 
quantum numbers of fields under the additional symmetries beyond the SM 
group. For the SM example in table \ref{tabpsSM}, the relevant SM Yukawa 
couplings are  $Q_L\cdot H_U\cdot U$, $Q_L \cdot H_D \cdot D$ and $L 
\cdot E\cdot H_U^* $.

In any event, the physical Yukawa couplings will also have a dependence 
on the corresponding K\"ahler potential, which in our models is less 
understood than in standard compactifications. Furthermore, in our models 
with anti D3-branes, the effective Yukawa couplings will have important 
supersymmetry breaking contributions. These could be useful in yielding a 
realistic pattern for the spectrum of fermion masses.

\item
{\it Soft Breaking Terms}

In the above discussion we have ignored the effect of fluxes on the 
world-volume gauge theory, which can be quite important. In principle, new 
couplings are generated from the interaction of the open string modes with 
the background flux, with a typical scale provided by the flux density at 
the bottom of the throat, namely the local string scale. The lowest 
dimension operators can be extracted by following the techniques developed 
recently in \cite{grana,ciu}, and provide scalar mass terms, gaugino 
masses, and scalar trilinear terms for the world-volume fields. 

We should distinguish diverse situations. In models with D3-branes, the fluxes
do not lead to any soft terms for 33 fields, but may provide additional 
couplings involving the 37 fields (e.g. the mass terms mentioned in 
section \ref{modustab}). The only source of soft terms breaking the local 
supersymmetric structure must be provided by the distant sources, like 
antibranes away from the throat, and hence their structure is very model 
dependent.

A more explicit description can be provided for models with anti 
D3-branes. As discussed in \cite{ciu}, the $\bar{3}\bar{3}$ fields acquire 
non-vanishing soft terms, related to the tensor structure of the flux at 
the location of that brane. On the other hand, the structure of flux-induced 
operators involving $\bar{3}7$ fields has not been yet determined. In any 
event, the scale of these couplings is expected to be the flux density, 
so it is of the order of the local fundamental scale, namely the TeV scale. 
These constructions hence have the potential to lead to a fully 
phenomenologically viable set of mass terms for fields beyond the Standard 
Model, but a full determination of their spectrum would require explicit 
expressions for the metric and flux backgrounds in our construction.

\end{enumerate}

\section{Discussion}

We have succeeded in providing an explicit construction that allows the 
introduction of realistic models on warped throats. The main idea in our 
construction is to combine configurations of branes at singularities (a 
very successful setup in the realisation of realistic chiral models with 
three quark-lepton families), with the construction of warped throats via 
fluxes (a successful setup in generating exponential hierarchies, fixing 
the moduli, obtaining de Sitter space, etc). This combines many of 
the successful ideas proposed during the past few years for the 
brane-world scenario, including a natural hierarchy, moduli stabilisation, 
supersymmetry breaking, etc. We emphasise that these are the first 
semirealistic models realising the RS generation of the electroweak/Planck 
hierarchy in string theory. Our construction then inherits many of the 
phenomenological properties of the Randall-Sundrum scenario and of models 
of D-branes at singularities, and has the potential to fix all moduli and 
obtain de Sitter space solutions. Given these promising features, a 
detailed phenomenological analysis of some of these models would be welcome. 
Some of these properties presumably require a more explicit description of 
the final metric and flux background in our construction, so further 
theoretical work on this issue is desirable.

One remarkable feature of our models is that the construction naturally 
introduces a $\IT^4$ at the bottom of the throat. This is convenient, in 
that it allows the introduction of D7-branes, which help in the 
construction of models with realistic spectrum.

There are many possible avenues to continue our work.
A complete  phenomenological analysis of some of our models would be
desirable in order to set the conditions under which some of the
models could be viable, issues of soft supersymmetry breaking terms,
Yukawa couplings, gauge unification are more relevant in our models
given that there is little room to tune parameters after the moduli
are fixed. Generic properties, such as the appearance of extra $Z'$
bosons at relatively low energy could be interesting to investigate
along the lines of \cite{giiq}. 
 Also, in the 
explicit construction we found that the $\IZ_3$ singularity was natural, 
as a possible symmetry of elliptic fibers in section 4, and convenient, 
because of the phenomenological virtues of D-branes at the $\IZ_3$ 
singularity. It would however be interesting to explore the possible
construction of warped throats with other orbifold (or non-orbifold) 
singularities.

There is a second mechanism to generate chiral models from D-branes, 
namely the intersecting brane models. It would be interesting to explore 
if the realistic models constructed in that approach could be also 
embedded in a warped throat background. In this sense, the explicit local models models built in \cite{uralocal} would be quite useful. Similarly, it would be  
interesting to explore the possibility of generalising the construction of 
semirealistic chiral models of magnetised D-branes with homogeneously 
distributed fluxes in \cite{blt,cascur}, to models with localised fluxes 
leading to strongly warped throats.

In reference \cite{dkkls} a new source of hierarchy was proposed in terms 
of tunnelling between multiple different throats similar to the ones 
discussed here, having several interesting potential implications.
It may be interesting to extend our models to multi-throats and realise 
this effect explicitly.

\medskip

The Klebanov-Strassler geometry was motivated from the gauge/gravity 
correspondence, as the gravitational dual of a non-conformal gauge 
theory. The dynamically generated chiral symmetry breaking infrared scale 
in the gauge theory is reproduced, in the gravity side, by the capped off 
throat. It would be interesting to better understand our constructions in 
that context, and provide a holographic gauge theory description for our 
more complicated structure of fluxes and cycles at the bottom of the 
throat. Notice that unlike Klebanov-Strassler, we do not have an explicit 
form of the metric defining the geometries, but certain gauge theory 
information may be obtained from the topological features of the model.
\footnote{Notice that the holographic interpretation of warped throats 
has been exploited in RS phenomenology discussions. We would like to 
point out a misconception in the related literature: it is often stated 
that the presence of SM fields in the infrared region of the throat 
implies that SM fields are dynamically generated bound states of the dual 
gauge theory. This is not the case in our constructions where SM fields 
can be introduced consistently in a way unrelated to the structure of the 
throat, and hence of the dual gauge theory side. In other words, the SM 
fields are present in the gravity side of the correspondence, not on the 
gauge theory side.}

Finally, we would like to mention the possible application of our formalism 
to provide concrete realistic examples of the recent attempts to obtain 
inflation from D-brane/anti-brane interactions in warped 
compactifications in string theory \cite{dbraneinflation,kklmmt,renata}. 

We expect much progress in these and other applications of the present models.

\section{Acknowledgements}
We thank R. Blumenhagen, C. Burgess, P. G. C\'amara, G. Honecker, 
L. Ib\'a\~nez and C. de la Roza for useful conversations. 
J.G.C. thanks M.P\'erez for her patience and affection. A.M.U. thanks 
DAMTP, Cambridge for hospitality when this project started, and M. 
Gonz\'alez for encouragement and support. 
J.G.C. is supported by the Ministerio de Educaci\'on, Cultura y Deporte 
through a FPU grant. F.Q. is partially funded by PPARC and the 
Royal Society Wolfson award. M.P.G.M. is supported by a postdoctoral 
grant of the Consejer\'{\i}a de Educaci\'on, Cultura, Juventud y 
Deportes de la Comunidad Aut\'onoma de La Rioja (Spain). Research by 
A.M.U. is partially supported by CICYT, Spain.

\appendix

\section{Homology relation}

In this appendix we prove the homology relation (\ref{homrelt}) by 
using geometric arguments. We will mainly follow the work of \cite{uralocal} 
and references therein. Let us first summarise the required tools. 

Remember that our geometry consists of a double elliptic fibration over a 
complex plane parameterised by $z$. Each elliptic fibration degenerates at 
three points related by the $\IZ_3$ symmetry, each labelled by the $(p,q)$ 
1-cycle that collapses on it. In our case, those 1-cycles are $(2,-1)$, 
$(-1,2)$, $(-1,-1)$ in each fibration. In order to distinguish the two 
elliptic fibrations, we adopt bare and primed labels for cycles and  
degeneration points, and arbitrarily call first(second) elliptic 
fibration to the bare(primed) one.

\begin{figure}
\begin{center}
\centering
\epsfysize=2.9cm
\leavevmode
\epsfbox{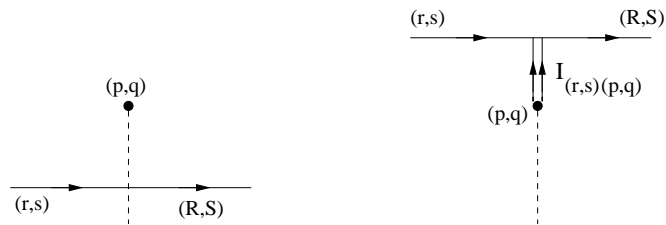}
\end{center}
\caption[]{\small Prong creation process.}
\label{prongcreat}
\end{figure}

In going around a degeneration fiber, the 1-cycles of the elliptic 
fibration suffer a $SL(2,\IZ)$ monodromy. A way to represent it is by 
considering a branch-cut in the $z$-plane outgoing each degeneration 
point. As a result, whenever one crosses (counterclockwise) the branch 
cut associated to a $(p,q)$ singular point of a given elliptic fibration, 
the $(r,s)$ 1-cycles in the corresponding elliptic fiber suffer a 
monodromy and turn into: 
\beqa
(r,s) \to (R,S)=(r,s)+I_{(r,s)(p,q)}(p,q)
\eeqa
where we have defined $I_{(r,s)(p,q)}\equiv rq-sp$ the intersection 
number of both 1-cycles in the fiber. This is depicted in figure 
\ref{prongcreat}a.

\begin{figure}
\begin{center}
\centering
\epsfysize=8.3cm
\leavevmode
\epsfbox{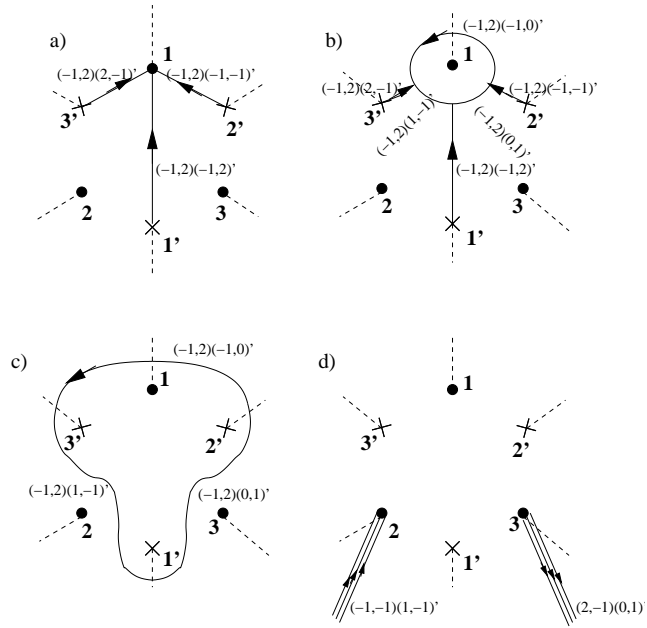}
\end{center}
\caption[]{\small Deformation argument: we draw the 1-cycle in the base 
($z$-plane) and, over each segment, specify the 2-cycle fibered over it. 
We describe the latter by two entries, each labelling the 1-cycle at each 
of the two elliptic fibrations. The notation for the singular points is \textbf{1}$=(-1,2)$, \textbf{2}$=(-1,-1)$, \textbf{3}$=(2,-1)$ (and primed), while the fibered 2-cycles are written explicitly.}
\label{deformation}
\end{figure}

Even though the location of the above branch-cuts is completely 
unphysical, their distribution does change the $(p,q)$ labels of the 
degeneration points. We fix the ambiguity by choosing our geometry as 
shown in figure \ref{deformation}a.

An important property of the structure of degenerations in our elliptic 
fibration is that the 1-cycle $(1,0)$ is invariant under the overall 
monodromy due to the three branch-cuts of the corresponding elliptic 
fiber. 

Besides the compact 3-cycles considered in the main text (a segment in 
the base starting and ending on a degeneration point with a 2-cycle 
fibered over it), one may also build up ``junction'' compact 3-cycles. 
These are cycles obtained by considering networks of oriented segments in 
the base (for which only the external legs must end up at a degeneration 
point) and fibering over it the appropriate 2-cycle in the double 
elliptic fibration. In order to obtain a consistent 3-cycle, the 
fibration must be such that i) homology charge of the fibered 2-cycle is 
conserved at each junction and ii) segments ending on a $(p,q)$/$(p,q)'$ 
degeneration point must have the corresponding $(p,q)$/$(p',q')$ cycle 
fibered (in order for the 3-cycle to be closed).

The existence of junction 3-cycles can be derived from the prong creation 
process, which we will use in our computation and briefly sketch in what 
follows \footnote{We focus on the case of just one elliptic fibration. 
This is the most general case for us, since we do not consider 
degeneration points common to both elliptic fibrations. At each 
collapsing point/branch cut of one of the elliptic fibrations, we can 
completely factorise the other.}. Consider a $(r,s)$ cycle going through a 
$(p,q)$ branch cut and thus becoming $(R,S)$. If we move the segment 
across the $(p,q)$ degeneration point, a number of prongs grow from the 
latter, and the original 3-cycle becomes a junction 3-cycle. The number 
of prongs outgoing the degeneration point is $I_{(r,s)(p,q)}$ (as 
required by conservation of the 2-homology charge at the junction) and 
necessarily have the $(p,q)$ 1-cycle fibered over it. The process is 
depicted in figure \ref{prongcreat}.

We are now ready to go through the deformation argument leading to the 
homology relation (\ref{homrelt}). The different steps of the process are 
displayed in figure \ref{deformation}. For the sake of clarity in the 
pictures, we recall the shorthand notation already introduced in the main 
text, namely denote the different degeneration points by \textbf{1,2,3}
$\equiv (-1,2),(-1,-1),(2,-1)$ (and the corresponding primed labels for 
the second elliptic fiber).

Our starting point is the ``arrow'' 3-cycle of figure \ref{deformation}a. 
According to the notation introduced in section \ref{fluxesthroat}, it is 
representative of the homology class $-([\Pi_{11'}]+[\Pi_{12'}]+[\Pi_{13'}])$. 
We now grow  a circle surrounding the \textbf{1} degeneration point, with 
the \textbf{1}$\equiv (-1,2)$ 1-cycle over the circle, so that no prong emanates at this step.
The situation looks like in figure \ref{deformation}b. Still we must choose 
a 1-cycle in the second elliptic fiber. Any choice would make the junction 
3-cycle consistent, but we choose it to be $(-1,0)'$ so that the prongs 
joining it to the primed degeneration points disappear at a later stage.

We now deform the cycle in the complex $z$-plane, crossing the three 
degeneration points of the primed elliptic fiber. By construction the 
prongs disappear and no new prongs are created, so that we are left 
with the situation depicted in figure \ref{deformation}c.

\begin{figure}
\begin{center}
\centering
\epsfysize=2.0cm
\leavevmode
\epsfbox{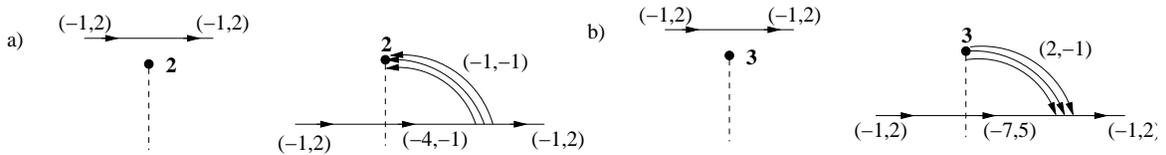}
\end{center}
\caption[]{\small This is the prong creation process taking place in the 
last step of our deformation argument. For instance, in a), once we cross the degeneration point, the $\textbf{1}\equiv (-1,2)$ 1-cycle  in the first elliptic fibration suffers a monodromie when going through the branch-cut and becomes $(-1,2)+3\times(-1,-1)=(-4,-1)$. Then 3 prongs incoming the \textbf{2}$\equiv(-1,-1)$ degeneration point emanate from the cycle and the original fibered 1-cycle $(-1,2)$ is recovered.}
\label{2details}
\end{figure}

The final step is to deform the 3-cycle so that it crosses the last two 
degeneration points of the first elliptic fiber, namely \textbf{2,3}. The 
procedure is detailed in figure \ref{2details}a and \ref{2details}b, with the 
outcome of 3 prongs incoming \textbf{2} and three outgoing \textbf{3}. If 
we finally stretch away the closed curve to the infinity of the base, we 
are left with the final non-compact homology 3-class 
$3[\Sigma_2]-3[\Sigma_3]$, thus proving the relation (\ref{homrelt}).

A final remark is in order. The homology relation we have obtained is 
valid up to a closed 3-cycle at infinity in the $z$-plane. In case we 
work with a non-compact space (equivalently, a non-compact base manifold), we 
can always push that cycle away to infinity and forget about it. However, 
when eventually embedding our setup into a compact space, the above 
homological relation should be considered just up to the closed cycle.  

One can use these geometric arguments to explore other homology relations 
among the cycles we use in the main paper. The only relevant ones for 
our discussion are the above, and its $\IZ_3$ related.  

\clearpage

\newpage

\section{Complete spectra of different models in section 5}

\begin{table}[!h!t] \footnotesize
\renewcommand{\arraystretch}{1.25}
\begin{center}
\begin{tabular}{|c|c|c|c|c|c|c|}
\hline Matter fields  &  $Q_3$  & $Q_L $ & $Q_R $ & $Q_{D7}$ & 
$Q_{{\ov {D3}}^i}$ & $B-L$   \\
\hline\hline ${{\bf \bar{3}\bar{3}}}$ \,\, $\NN=1$ Ch. Mults. 
 &  & & & & & \\
\hline $3(3,2,1;1)$ & 1  & -1 & 0 & 0 & 0 & 1/3  \\
\hline $3(\bar 3,1,2;1)$ & -1  & 0  & 1 & 0 & 0 & -1/3 \\
\hline $3(1,2,2;1)$ & 0  & 1  & -1 & 0 & 0 & 0  \\
\hline\hline ${\bf {\bar 3}7}$ \,\, Ch. Fermions & & & & & & \\
\hline $(1,2,1;{\bar 3})$ & 0 & 1 & 0 & -1  & 0 & -1 \\
\hline $(1,1,2;3)$ & 0 & 0 & -1 & 1 & 0  & 1 \\
\hline\hline ${\bf {\bar 3}7}$ \,\, Cmplx.Scalars & & & & & & \\
\hline $(3,1,1;{\bar 3})$ & 1 & 0 & 0 & -1  & 0 & -2/3 \\
\hline $({\bar 3},1,1;3)$ & -1 & 0 & 0 & 1 & 0  & 2/3 \\
\hline\hline ${\bf {\bar 3}_i7}$ \,\, Cmplx.Scalars & & & & & & \\
\hline $(1,1,1;{\bar 3})$ & 0 & 0 & 0 & -1  & $1_i$ & 0\\
\hline $(1,1,1;3)$ & 0 & 0 & 0 & 1 & -$1_i$  & 0 \\
\hline 
\end{tabular}
\end{center} \caption{Spectrum of $SU(3)\times SU(2)_L\times SU(2)_R$
model. We present the quantum numbers  under the $U(1)$ groups. The
first three $U(1)$'s arise from the $\ov{\rm{D3}}$-brane sector at the 
origin. The next two come from the D7- and additional $\ov{\rm{D3}}$-brane 
sectors, and are written as a single column,  distinguished with a 
label $i$.
\label{tabpslr} }
\end{table}     

\begin{table}[!h!t] \footnotesize
\renewcommand{\arraystretch}{1.25}
\begin{center}
\begin{tabular}{|c|c|c|c|c|c|c|c|c|c|}
\hline 
Matter fields  &  $Q_3$  & $Q_2 $ & $Q_1 $ & $Q_{D7;U(6)}$ & 
$Q_{D7;U(3)}$ & $Q_{D3;U(2),i}$ & $Q_{D3;U(1),i}$ & $Y$ & Name   \\
\hline\hline ${{\bf \bar{3}\bar{3}}}$\,\, $\NN=1$ Ch.\, Mults. &  & & & & 
& & & & \\
\hline $3(3,\bar{2};1,1;1)$ & 1  & -1 & 0 & 0 & 0 & 0 & 0 & 1/6 & $Q_L$  \\
\hline $3(1,2;1,1;1)$ & 0  & 1  &-1 & 0 & 0 & 0 & 0 & 1/2 & $H_U$ \\
\hline $3(\bar{3},1;1,1;1)$ & -1  & 0  & 1 & 0 & 0 & 0 & 0 & -2/3 & $U$ \\
\hline\hline ${\bf {\bar 3}7}$  Ch.Ferms. & & & & & & & & & \\
\hline $(\bar 3,1;1,3;1)$ & -1 & 0 & 0 & 0  & 1 & 0 & 0 & 1/3 & $D$ \\
\hline $(1,1;6,1;1)$ & 0 & 0 & -1 & 1  & 0 & 0 & 0 & 1 & $E$ \\
\hline $(1,2,;\bar{6},1;1)$ & 0 & 1 & 0 & -1 & 0  & 0 & 0 & -1/2 & $L$ \\
\hline $(1,1;1,\bar 3;1)$ & 0 & 0 & 1 & 0 & -1  & 0 & 0 &-1 & $\bar{E}$ \\
\hline\hline ${\bf {\bar 3}7}$ \,\, Cmplx.Scalars & & & & & & & & & \\
\hline $(\bar 3,1;6,1;1)$ & -1 & 0 & 0 & 1  & 0 & 0 & 0 & 1/3 & \\
\hline $(3,1;\bar 6,1;1)$ & 1 & 0 & 0 & -1  & 0 & 0 & 0 & -1/3 & \\
\hline $(1,2,;1,\bar{3};1)$ & 0 & 1 & 0 & 0 & -1  & 0 & 0 & -1/2 & $H_D$\\
\hline $(1,2,;1,3;1)$ & 0 & -1 & 0 & 0 & 1  & 0 & 0 & 1/2 & \\
\hline\hline ${\bf {\bar 3}_i7}$ \,\, Ch.Ferms. & & & & & & & & \\
\hline $(1,1;1,3;2_i)$ & 0 & 0 & 0 & 0  & 1 & -1$_i$ & 0 & 0 & \\
\hline $(1,1;\bar{6},1;1)$ & 0 & 0 & 0 & -1 & 0  & 0 & 1$_i$ & 0 & \\
\hline\hline ${\bf {\bar 3}_i7}$ \,\, Cmplx.Scalars & & & & & & &  & \\
\hline $(1,1;6,1;2_i)$ & 0 & 0 & 0 & 1  & 0 & -1$_i$ & 0 & 0 & \\
\hline $(1,1;\bar 6,1;2_i)$ & 0 & 0 & 0 & -1  & 0 & 1$_i$ & 0 & 0 & \\
\hline $(1,1;1,\bar{3};1)$ & 0 & 0 & 0 & 0 & -1  & 0 & 1$_i$ & 0 & \\
\hline $(1,1;1,3;1)$ & 0 & 0 & 0 & 0 & 1  & 0 & -1$_i$ & 0 & \\
\hline\hline ${\bf 77}$ $\NN=1'$ Ch.Mult. & & & & & & & & & \\
\hline $3(1,1;6,\bar{3};1)$ & 0 & 0 & 0 & 1  & -1 & 0 & 0 & 0 &  \\
\hline 
\end{tabular}
\end{center}
\caption[]{Spectrum of the $SU(3)\times SU(2)\times U(1)$
Standard Model. We present the quantum numbers  under the $SU(3) \times 
SU(2)$ on $\ov {\rm D3}$-branes at the origin, the $SU(6)\times SU(3)$ on 
D7-branes, the $SU(2)_i$ on $\ov {\rm D3}$-branes at the $i^{th}$ fixed 
point away from the origin, and also under the $U(1)$ factors. The
first three $U(1)$'s arise from the $\ov{\rm{D3}}$-brane sector at the 
origin, the next two come from the D7-brane sectors, and the last two 
from $\ov {\rm D3}$-branes away from the origin.}
\label{tabpsSM} 
\end{table}

\begin{table}[htb] \footnotesize
\renewcommand{\arraystretch}{1.25}
\begin{center}
\begin{tabular}{|c|c|c|c|c|c|c|c|c|}
\hline 
Matter fields  &  $Q_3$  & $Q_2 $ & $Q_1 $ & $Q_{D7;U(3)}$ & 
$Q_{D7;U(6)}$ & $Q_{D3;U(2),i}$ & $Q_{D3;U(1),i}$ & $Y$   \\
\hline\hline ${{\bf 33}}$\,\, $\NN=1$ Ch.\, Mults. &  & & & & 
& & &\\
\hline $3(3,\bar{2};1,1;1)$ & 1  & -1 & 0 & 0 & 0 & 0 & 0 & 1/6  \\
\hline $3(1,2;1,1;1)$ & 0  & 1  &-1 & 0 & 0 & 0 & 0 & 1/2 \\
\hline $3(\bar{3},1;1,1;1)$ & -1  & 0  & 1 & 0 & 0 & 0 & 0 & -2/3  \\
\hline\hline ${\bf 37+73}$  $\NN=1$ Ch.\, Mults. & & & & & & & & \\
\hline $(3,1;\bar 3,1;1)$ & 1 & 0 & 0 & -1  & 0 & 0 & 0 & -1/3 \\
\hline $(1,2;1,\bar 6;1)$ & 0 & 1 & 0 & 0  & -1 & 0 & 0 & -1/2 \\
\hline $(1,1;3,1;1)$ & 0 & 0 & -1 & 1 & 0  & 0 & 0 & 1\\
\hline $(\bar 3,1;1,6;1)$ & -1 & 0 & 0 & 0 & 1  & 0 & 0 & 1/3 \\
\hline\hline ${\bf 3_i7+73_i}$ $\NN=1$ Ch.\, Mults. & & & & & & & \\
\hline $(1,1;\bar 3,1;2_i)$ & 0 & 0 & 0 & -1  & 0 & 1$_i$ & 0 & 0 \\
\hline $(1,1;1,\bar{6};1_i)$ & 0 & 0 & 0 & 0 & -1  & 0 & 1$_i$ & 0 \\
\hline $(1,1;1,6;\bar 2_i)$ & 0 & 0 & 0 & 0 & 1  & -1$_i$ & 0 & 0 \\
\hline\hline ${\bf 77}$ $\NN=1$ Ch.Mult. & & & & & & & &\\
\hline $3(1,1;3,\bar{6};1)$ & 0 & 0 & 0 & 1  & -1 & 0 & 0 & 0 \\
\hline 
\end{tabular}
\end{center}
\caption[]{Spectrum of the $SU(3)\times SU(2)\times U(1)$
Standard Model. We present the quantum numbers  under the $SU(3) \times 
SU(2)$ on ${\rm D3}$-branes at the origin, the $SU(3)\times SU(6)$ on 
D7-branes, the $SU(2)_i$ on ${\rm D3}$-branes at the $i^{th}$ fixed 
point away from the origin, and also under the $U(1)$ factors. The
first three $U(1)$'s arise from the $\textrm{D3}$-brane sector at the 
origin, the next two come from the D7-brane sectors, and the last two 
from ${\rm D3}$-branes away from the origin.}
\label{tabpssm1} 
\end{table}

\begin{table}[!h!t] \footnotesize
\renewcommand{\arraystretch}{1.25}
\begin{center}
\begin{tabular}{|c|c|c|c|c|c|c|}
\hline Matter fields  &  $Q_4$  & $Q_L $ & $Q_R $ & $Q_{D7}$ & 
$Q_{{\ov {D3}}^i}$ & $Q$   \\
\hline\hline ${{\bf \bar{3}\bar{3}}}$ \,\, $\NN=1$ Ch. Mults. 
 &  & & & & & \\
\hline $3(4, 2,1;1;1)$ & 1  & -1 & 0 & 0 & 0 & -1/4  \\
\hline $3(\bar 4,1,2;1;1)$ & -1  & 0  & 1 & 0 & 0 & 1/4 \\
\hline $3(1,2,2;1;1)$ & 0  & 1  & -1 & 0 & 0 & 0  \\
\hline\hline ${\bf {\bar 3}7}$ \,\, Ch. Fermions & & & & & & \\
\hline $(1,2,1;{\bar 6};1)$ & 0 & 1 & 0 & -1  & 0 & 1/2 \\
\hline $(1,1,2;6;1)$ & 0 & 0 & -1 & 1 & 0  & -1/2 \\
\hline\hline ${\bf {\bar 3}7}$ \,\, Cmplx.Scalars & & & & & & \\
\hline $(4,1,1;{\bar 6};1)$ & 1 & 0 & 0 & -1  & 0 & 1/4 \\
\hline $({\bar 4},1,2;6;1)$ & -1 & 0 & 0 & 1 & 0  & -1/4 \\
\hline\hline ${\bf {\bar 3}_i7}$ \,\, Cmplx.Scalars & & & & & & \\
\hline $(1,1,1;{\bar 6};2_i)$ & 0 & 0 & 0 & -1  & $1_i$ & 0\\
\hline $(1,1,1;6;2_i)$ & 0 & 0 & 0 & 1 & -$1_i$  & 0 \\
\hline \end{tabular}
\end{center} \caption{Spectrum of $SU(4)\times SU(2)_L\times SU(2)_R$
model. We present the quantum numbers under the $SU(4)\times SU(2)\times 
SU(2)$ on $\ov{\rm D3}$-branes at the origin, the $SU(6)$ on D7-branes, 
and the $SU(2)_i$ on $\ov{\rm D3}$-branes away from the origin, and the 
different $U(1)$ groups. The first three $U(1)$'s arise from the 
$\ov{\rm{D3}}$-brane sector at the origin. The next two come from the 
D7-branes and the additional $\ov{\rm{D3}}$-brane sectors, with the latter 
written as a single column, distinguished with a label $i$.
\label{tabps3r}}
\end{table}

\end{document}